\documentclass[iicol,sn-mathphys-num]{sn-jnl}
\usepackage{graphicx}%
\usepackage{multirow}%
\usepackage{amsmath,amssymb,amsfonts}%
\usepackage{amsthm}%
\usepackage{mathrsfs}%
\usepackage[title]{appendix}%
\usepackage{xcolor}%
\usepackage{textcomp}%
\usepackage{manyfoot}%
\usepackage{booktabs}%
\usepackage{algorithm}%
\usepackage{algorithmicx}%
\usepackage{algpseudocode}%
\usepackage{listings}%
\usepackage{subcaption}
\usepackage{capt-of}

\theoremstyle{thmstyleone}

\theoremstyle{thmstyletwo}

\theoremstyle{thmstylethree}

\begin{document}

\title[Article Title]{iTrace: Click‑Based Gaze Visualization on the Apple Vision Pro}

\author[1]{\fnm{Esra} \sur{Mehmedova}}\email{esra.mehmedova@tum.de}
\author*[1]{\fnm{Santiago} \sur{Berrezueta-Guzman}}\email{s.berrezueta@tum.de}
\author[1]{\fnm{Stefan} \sur{Wagner}}\email{stefan.wagner@tum.de}

\affil*[1]{\orgdiv{Chair of Software Engineering}, \orgname{Technical University of Munich}, \orgaddress{\street{Bildungscampus 2}, \city{Heilbronn}, \country{Germany}}}

\abstract{

The Apple Vision Pro is equipped with accurate eye-tracking capabilities; however, the device's privacy restrictions prevent direct access to continuous user gaze data. This study introduces \textit{iTrace}. This novel application overcomes these limitations through click-based gaze extraction techniques, including manual methods such as a pinch gesture, as well as automatic approaches that utilize dwell control or a gaming controller. We developed a system with a client-server architecture that captures the gaze coordinates and transforms them into dynamic heatmaps for video and spatial eye tracking. The system can generate individual and averaged heatmaps, enabling analysis of personal and collective attention patterns. 

To demonstrate its effectiveness and evaluate the usability and performance, a study was conducted with two groups of 10 participants, each testing different clicking methods. The 8BitDo controller achieved higher average data collection rates at 14.22 clicks/s compared to 0.45 clicks/s with dwell control, enabling significantly denser heatmap visualizations. The resulting heatmaps reveal distinct attention patterns, including concentrated focus in lecture videos and broader scanning during problem-solving tasks. By allowing dynamic attention visualization while maintaining a high gaze precision of 91\%, iTrace demonstrates strong potential for a wide range of applications, including educational content engagement, environmental design evaluation, marketing analysis, and clinical cognitive assessment. Despite the current gaze data restrictions on the Apple Vision Pro, we encourage developers to use \textit{iTrace} only in research settings.

}


\maketitle
\section{Introduction}\label{I}

Eye tracking refers to the method of recording gaze movements and locations over time \cite{hu2022eye, duchowski2017eye}. This information can provide valuable insights into user attention patterns and even reveal specific conscious and unconscious thought processes \cite{kroger2020does}. In recent years, eye tracking has gained significant popularity \cite{carter2020best}, primarily due to the advancements in tracking devices, which have become cheaper and more widely available \cite{wedel2017review}. Introducing new wearable devices further contributes to this trend by allowing users to move freely during data collection. This opens new opportunities and applications for gaze data collection in real-world environments \cite{cognolato2018head, bisogni2024gaze}.

The \textit{Apple Vision Pro} is a mixed reality headset equipped with multiple infrared cameras for accurately detecting the user's gaze \cite{cheng2024first}. Currently, the eye-tracking capabilities on this device are primarily used to enable interaction with UI elements \cite{kaye2024apple}. Notably, the eye-tracking accuracy of the Apple Vision Pro is comparable to, and in some cases better than, that of specific virtual reality (VR) devices and eye-tracking glasses, making it a promising tool for gaze-based analysis and visualization \cite{hu2024apple}. Due to privacy policies,  Apple Inc. does not provide developers with direct access to the continuous raw eye-tracking data, which presents significant challenges for extracting the gaze coordinates necessary for further research and analysis \cite{huang2024measuring}. Given this constraint, our work presents a novel approach for collecting and visualizing gaze data on the Apple Vision Pro.

This study addresses the limitations of accessing raw gaze coordinates by introducing a click-based method for extracting eye-tracking data. We propose and evaluate multiple clicking techniques, including a \textit{pinch gesture}, \textit{dwell control}, and a \textit{gaming controller}, each best suited for different research objectives and application contexts. Due to the wide range of domains in which eye tracking can be used, various approaches are available for visualizing and analyzing gaze data \cite{blascheck2017visualization}. This study presents the eye-tracking data as a dynamic heatmap, visualizing the users' attention patterns over time. Our system generates individual and average heat maps, enabling analysis of personal viewing behaviors and collective attention patterns across multiple users. 

To collect and visualize eye-tracking data on the Apple Vision Pro, we developed an application called \textit{iTrace}\footnote{The \textit{iTrace} application code is available on GitHub: \url{https://github.com/TUM-HN/iTrace}}. As shown in Figure~\ref{fig:eye_tracking}, the system consists of two main functionalities: \textbf{video eye tracking}, which captures gaze patterns. At the same time, users watch pre-recorded content, and \textbf{spatial eye tracking}, which records gaze behavior within real-world environments. Both functionalities are implemented through a client-server architecture that processes the collected gaze coordinates into a heatmap visualization. 

Our system provides researchers with valuable tools for directly analyzing visual attention in videos and real-world environments on the Apple Vision Pro. This work contributes to the growing field of eye tracking research by expanding the available tools for gaze analysis in mixed reality headsets. 

\section{Related Work}\label{RW}

Eye‐tracking research has leveraged devices ranging from webcams and glasses to modern head‐mounted displays (HMDs), each offering unique benefits and facing specific limitations. Various visualization techniques, such as heatmaps, scanpaths, and attention maps, have been developed to interpret gaze data in both 2D and 3D contexts.

\subsection{Devices Used for Eye Tracking}

A wide range of devices is used for eye-tracking purposes, each best suited to different research needs and application contexts. Kaduk et al. \cite{kaduk2024webcam} emphasized how integrating webcam-based eye-tracking systems has significantly expanded access to gaze research by utilizing commonly available devices such as personal computers, laptops, tablets, and mobile phones. These systems are limited to user mobility, but offer convenience and easy access. The study found that webcam eye trackers are generally less precise than specialized laboratory-grade systems, such as the EyeLink 100, a stationary eye-tracking device positioned near screen displays.

A different approach to detecting user gaze is the use of eye-tracking glasses. These wearable devices combine pupil and scene cameras to capture gaze within real-world environments. MacInnes et al. \cite{macinnes2018wearable} demonstrated that eye-tracking glasses, like the Pupil Labs, SMI ETG, and Tobii glasses, enable accurate gaze tracking while allowing users full mobility. They also highlighted the increased complexity of accurately mapping 2D gaze coordinates to targets in 3D environments, as opposed to flat-screen-based content.

With the recent advancements in eye-tracking technologies in HMDs, new possibilities for gaze-based analysis have become available \cite{cao2023mobile}. Unlike traditional eye-tracking glasses, HMDs like the Apple Vision Pro and HTC Vive Pro Eye enable eye tracking to be fully integrated with your virtual and immersive environments. Huang et al. \cite{huang2024measuring}, in their study "Measuring eye-tracking accuracy and its impact on usability in Apple Vision Pro", concluded that the Apple Vision Pro achieved better eye-tracking accuracy compared to the SMI ETG2 and Tobii Pro 2 glasses. This highlights the potential of using HMDs for eye-tracking analysis in both virtual and real-world environments.

This study utilizes the Apple Vision Pro to capture and visualize the user’s gaze, a feature that, to our knowledge, has not yet been implemented due to Apple’s current restrictions on accessing the raw eye-tracking data.

\subsection{Eye Tracking Visualization}

Depending on the application and use case, various approaches can be used to represent gaze data. Blascheck et al. \cite{blascheck2017visualization} reviewed 110 research papers to provide a comprehensive taxonomy of eye tracking visualization techniques. The research identified different types of gaze visualization, including statistical graphs, timeline visualizations, attention maps, scanpath visualizations, and more. Their survey also differentiates between 3 main approaches for 3D heatmap visualization: projected, object-based, and surface-based.

Web-based eye tracking is commonly used due to its easy accessibility and cost-effectiveness. It enables researchers to conduct usability studies on various devices without requiring specialized hardware. GazeRecorder is an eye-tracking software that generates dynamic heatmap videos by overlaying gaze data captured by a web camera onto screen recordings. These visualizations, however, often lack access to the gaze coordinate data used in the heatmap videos. 
Gnanaraj et al. \cite{gnanaraj2025point} aimed to enhance the utility of webcam-based eye-tracking systems by developing a deep learning-based solution to extract the gaze points from dynamic heatmap videos generated during eye-tracking sessions. The study also emphasized the importance of having access to the raw gaze data, along with its visual representation, which can facilitate further analysis of the user's gaze patterns.

Wearable eye-tracking glasses have also been used for gaze visualization in screen-based studies. For example, Chrześcijańska \cite{chrzescijanska2024properly} utilized Tobii eye-tracking glasses to record participants’ attention as they viewed floor plans displayed on a computer screen. The study's primary objective was to assess the influence of floor plan design formats and elements on property purchasing intentions. The study analyzed viewing duration, fixation duration, the number of fixations, and the number of saccades in 180 participants. The gaze data was then visualized as a heatmap projected over the floor plans. The participants were also asked to complete a questionnaire to understand how the different floor plan visualizations influenced their decision-making.

Beyond flat screens, gaze visualization can also be applied to immersive videos. Löwe et al. \cite{lowe2017gaze} developed a visual framework to analyze head and eye movements while watching an immersive 360° video. Their approach combined scanpath and heatmap visualizations to represent the gaze and head orientation data of the panoramic scene. The participants used HMDs to view the immersive video content, which enabled them to achieve those results.

A more challenging approach to visualizing eye tracking data is to translate it into three-dimensional spaces. Li et al. \cite{li2020visualization} utilized Tobii eye-tracking glasses to capture gaze data and video footage of the user's surroundings. Their primary objective was to develop a method for visualizing users' 3D gaze fixations on reconstructed models of real scenes. Using image-based 3D reconstruction, they created a 3D mesh of the environments from the recorded video frames. They then translated the 2D gaze data onto the 3D model. The resulting visualization presented a 3D heatmap overlaid on the reconstructed world mesh, accurately representing the user's gaze.

Gaze can also be visualized directly onto VR environments. Bianconi et al. \cite{bianconi2019immersive} explored wayfinding and spatial legibility within a VR reconstruction of a real campus building using the HTC Vive headset equipped with eye-tracking lenses. Their goal was to assess how users perceive and navigate complex architectural environments. To achieve this, they tracked the participants' gaze as they explored the immersive 3D space and then visualized their gaze patterns as 3D heatmaps embedded within the VR environment.

The current study achieves eye tracking visualization by implementing two features within the Apple Vision Pro's mixed reality environment. The first feature is eye tracking on video content, which brings mobility to previous screen-based eye tracking solutions by allowing users to view video content as a floating window. Their gaze is tracked and visualized as a heatmap or attention map overlaid on top of the video. The second feature implements spatial eye tracking, where a video recording of the user's surrounding environment is captured simultaneously while the user's gaze is tracked. The gaze data is then projected as a 2D heatmap directly onto the recorded video. Both features provide detailed eye-tracking data, including precise gaze coordinates in JSON format alongside the heatmap videos. This can enable further analysis and research applications beyond the visualization capabilities.

\section{Methodology}\label{M}

We designed a click-based system to capture and visualize gaze data on the Vision Pro, despite its restrictions on accessing gaze data. Through calibration tests and heatmap generation, we compare different interaction methods and analyze attention patterns in video and real‑world settings.

\subsection{System Architecture Overview}

\begin{figure*}[t] 
\centering
  \begin{subfigure}[b]{0.43\linewidth}
    \centering
    \includegraphics[width=\linewidth]{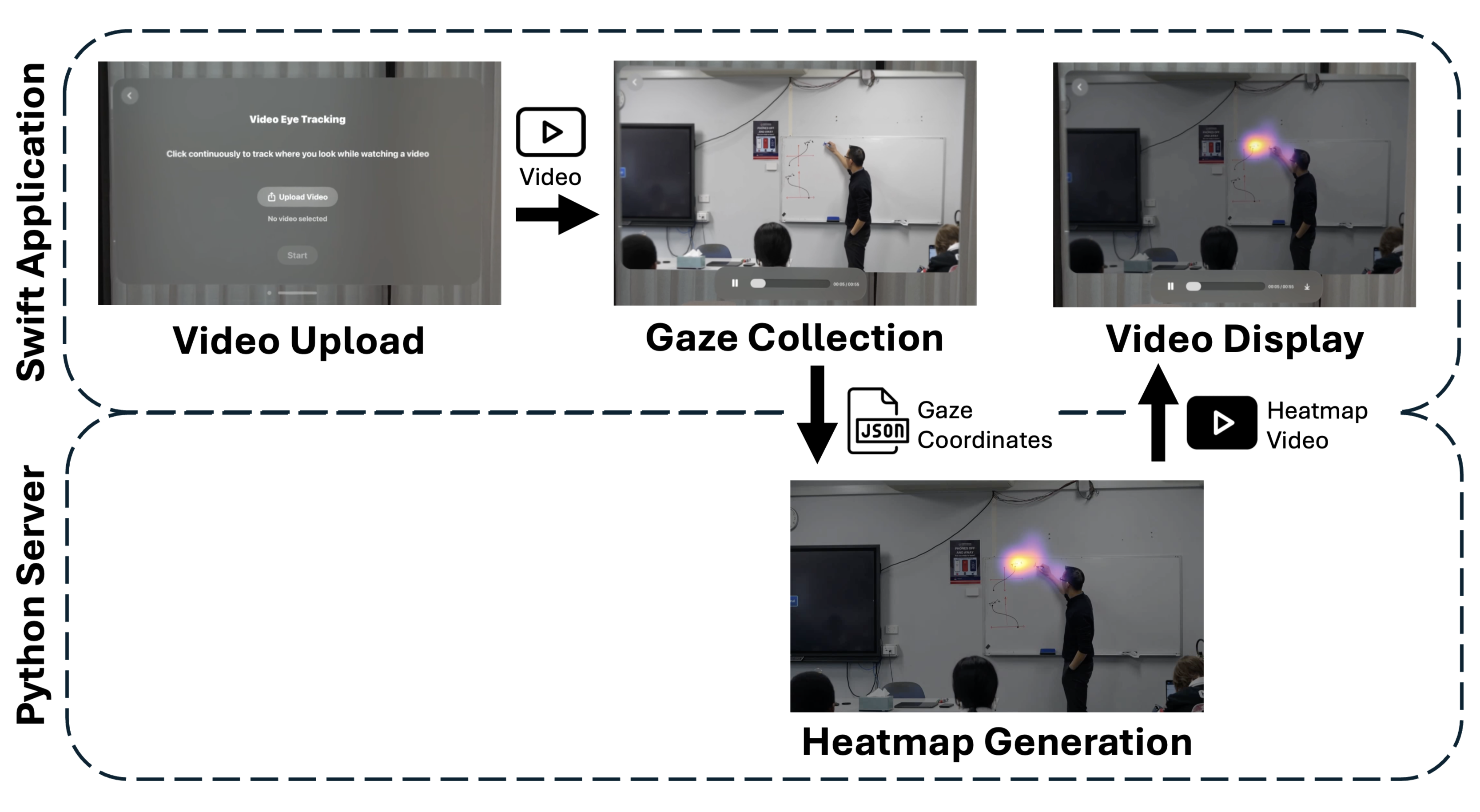}
  \end{subfigure}
  \hfill
  \begin{subfigure}[b]{0.56\linewidth}
    \centering
    \includegraphics[width=\linewidth]{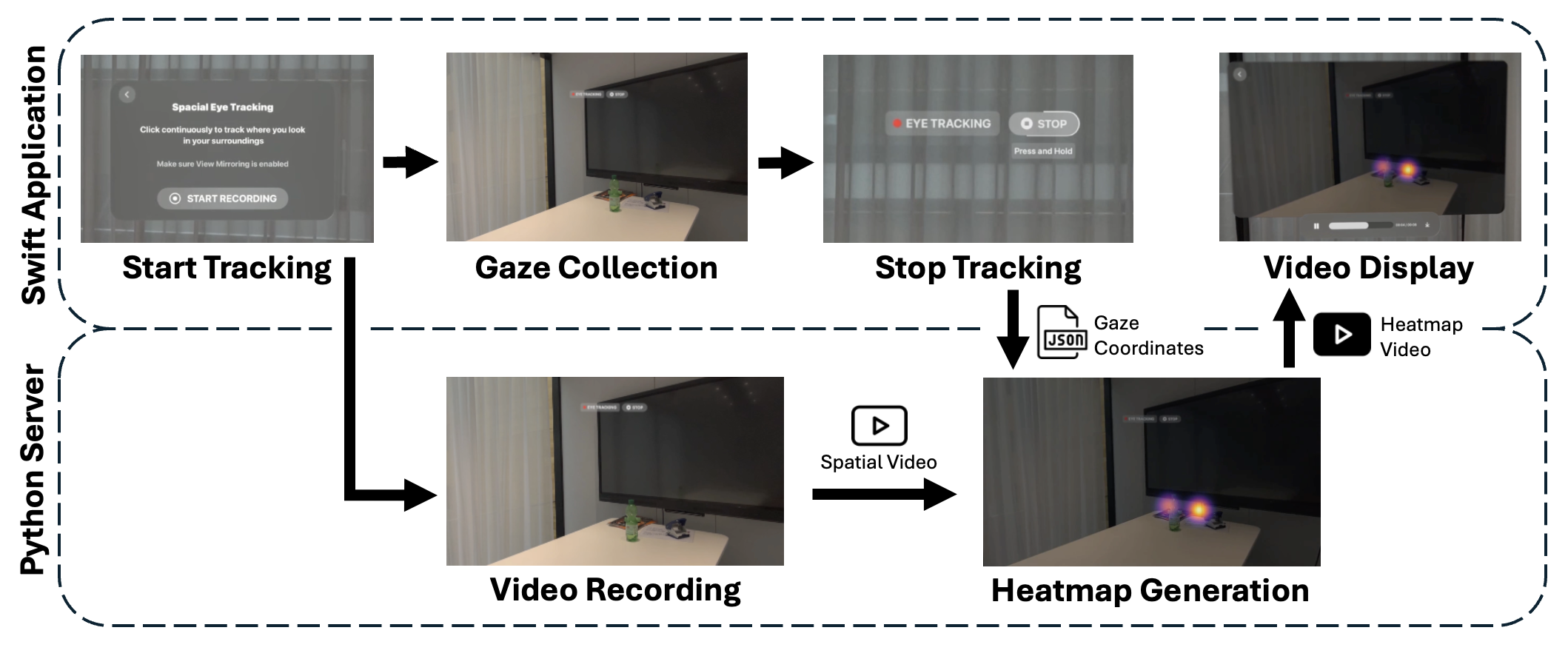}
  \end{subfigure}
  \caption{The \textit{iTrace} pipeline for click‑based gaze mapping on the Apple Vision Pro—(left) \textit{video eye tracking}: the Swift app captures and sends the gaze data to the server to produce heatmap videos; (right) \textit{spatial eye tracking}: the application triggers environment recording and gaze capture, then the server overlays heatmaps on the mirrored footage.}
  \label{fig:eye_tracking}
\end{figure*}

The \textit{iTrace} heatmap generation system, as illustrated in Figure~\ref{fig:eye_tracking}, consists of two main components operating in a client-server architecture: a Swift application on an Apple Vision Pro device and a server on a macOS device. Both components must be connected to the same local network to allow direct data transfer. The client-server communication utilizes HTTP protocols with a JSON data exchange format. The system also implements a zero-configuration networking approach using Zeroconf service discovery, allowing the Vision Pro application to automatically locate and connect to the Python server on the local network. This approach eliminates the need for manual IP address configuration, ensuring seamless connectivity between the client and server.

\subsubsection{Vision Pro Swift Application}

The client application is developed using Swift due to its native integration and optimization for the Apple Vision Pro platform. It serves as the primary user interface (UI) and data collection component. The application is responsible for presenting visual content to the user, capturing gaze coordinates, and transferring relevant data to the server for further processing.  

\subsubsection{Python Flask Server} 

The server backend is implemented using Python to leverage its extensive ecosystem of libraries for video processing, image manipulation, and data analysis. The server component handles computationally intensive tasks, including generating heatmaps and screen recordings. Built on the Flask web framework, the server provides RESTful API endpoints for communication with the Vision Pro client.

\subsection{Gaze Data Collection}

\textit{iTrace} overcomes the Apple Vision Pro's eye-tracking limitations and allows capturing gaze indirectly by logging the on-screen coordinates exposed at the moment of an intentional interaction. 

\subsubsection{Apple Vision Pro's Eye-Tracking Limitations}

The Apple Vision Pro, while equipped with advanced eye-tracking functionalities, restricts direct access to the raw gaze data. Apple's user data protection regulations prevent developers from accessing the continuous, real-time eye-tracking information without explicit user interaction \cite{apple2024visionproprivacy}. 
As a result, gaze information is only made available to applications when the user intentionally interacts with the content, typically by looking at an object and performing a finger tap gesture.

This privacy policy is detrimental to eye-tracking research as it limits access to the continuous gaze data necessary for detailed behavioral analysis. Unlike traditional eye-tracking systems that rely on high-frequency gaze sampling rates, gaze research on the Apple Vision Pro is restricted to specific user interaction events. Given those constraints, this research implemented a click-based gaze extraction approach to collect eye-tracking data. This methodology leverages the Vision Pro's ability to provide gaze coordinates during user interaction through various clicking methods. While this approach introduces limitations in data collection frequency compared to continuous tracking, it demonstrates the most viable solution within the current privacy restrictions, enabling meaningful analysis of gaze behavior.

\subsubsection{Click-Based Gaze Extraction Techniques}

\begin{figure}[htbp]
  \centering
  \begin{subfigure}[b]{0.53\linewidth}
    \centering
  \includegraphics[width=1\linewidth]{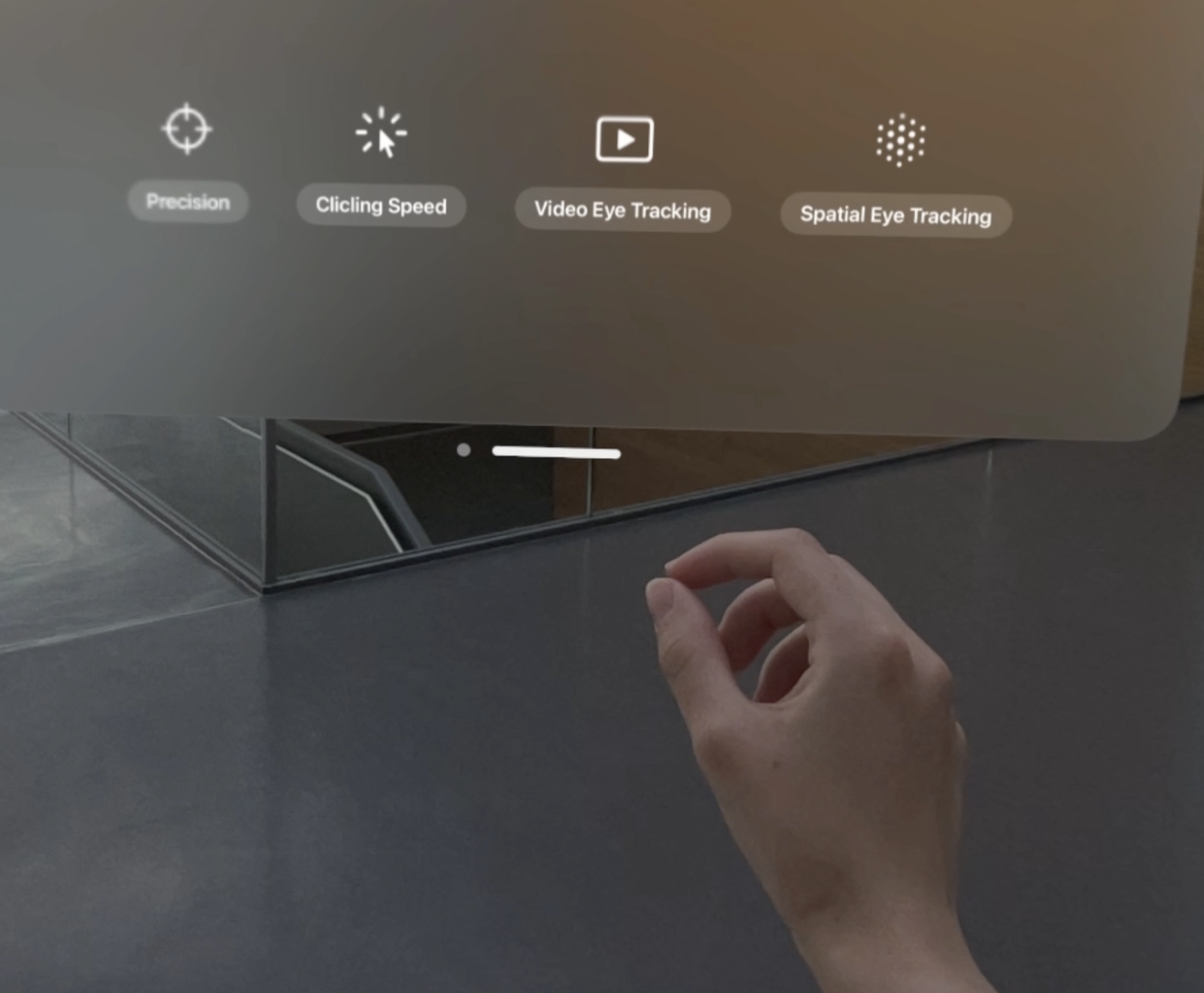}
  \end{subfigure}
  \hfill
  \begin{subfigure}[b]{0.53\linewidth}
    \centering
  \includegraphics[width=1\linewidth]{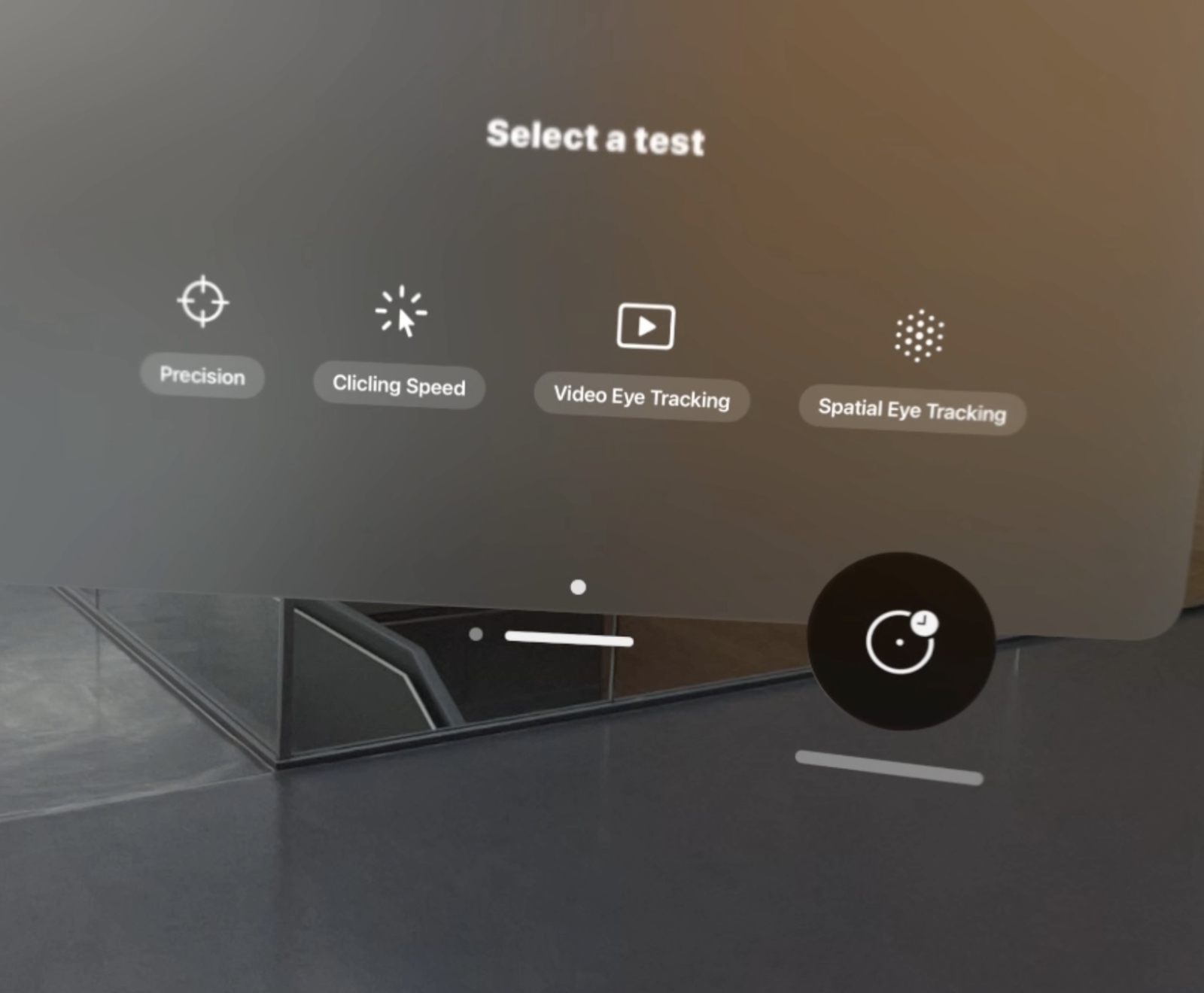}
  \end{subfigure}
  \hfill
  \begin{subfigure}[b]{0.53\linewidth}
    \centering
  \includegraphics[width=1\linewidth]{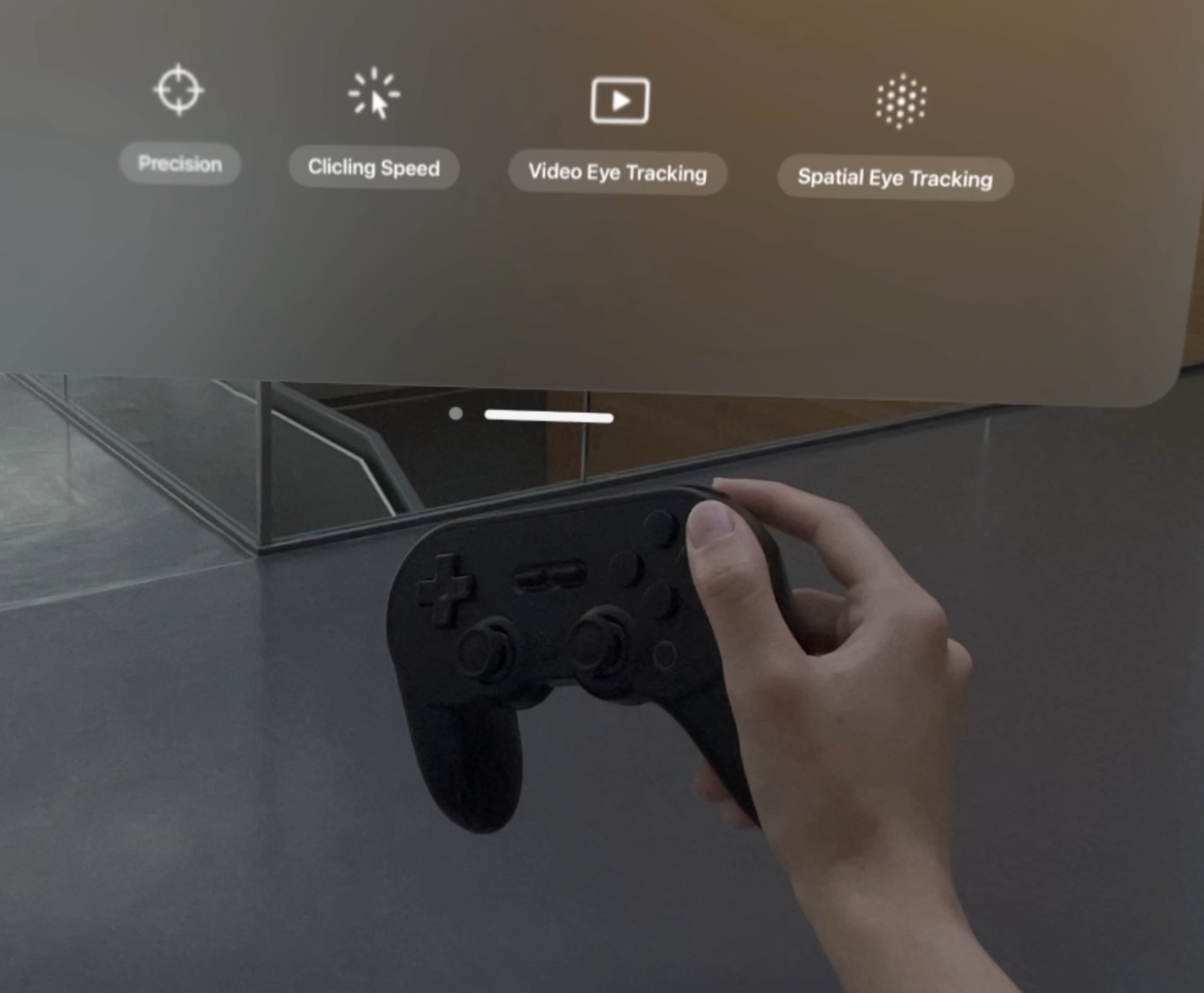}
  \end{subfigure}
  
  \caption{Click‐based interaction methods: (top) pinch gesture, (middle) dwell control, and (bottom) gaming controller.}
  \label{fig:clicking_methods}
\end{figure}

Three distinct clicking methods were explored, as demonstrated in Figure~\ref{fig:clicking_methods}. Each presents unique advantages and limitations regarding user comfort and data collection frequency.

\textbf{Manual pinch gesture} with your index finger and thumb is the most common way of clicking on the Vision Pro. It utilizes the device's built-in hand-tracking capabilities to detect tap gestures. While this method allows the user to perform on average \textit{6.8 clicks per second}, it also introduces significant user fatigue after extended periods of manual clicking. This makes the pinch gesture clicking approach unsuitable for long eye-tracking sessions.

\textbf{Dwell control} is an accessibility setting used to interact with the device hands-free. It allows users to select an element by simply keeping their gaze on it for a specified duration. When set to the fastest setting, dwell control can perform clicks every few seconds; however, the time interval between clicks is not fixed, as it depends on the user's gaze movement. The frequency of the clicks is influenced by the movement tolerance defined in the dwell control settings. With a higher tolerance, dwell control can allow users to perform clicks with a larger range of gaze movement. If the user's gaze moves too much and too often and exceeds the set tolerance, no clicks will be registered, and as a result, no gaze data will be collected. When the user's gaze remains steady within the allowed movement range, dwell control can perform \textit{0.7 clicks per second on average}. While this method introduces significant constraints on data collection frequency, it also eliminates physical fatigue issues associated with traditional methods.

\textbf{Gaming controllers} can also be utilized as an alternative to clicking. Bluetooth controllers supported by VisionOS, with embedded turbo-clicking functionalities, can perform fast repeated clicks with a single button press. When connected to the Apple Vision Pro, controllers like the 8BitDo Pro 2 can be customized to perform turbo clicking while the user holds a button. This approach allows for high-frequency data collection, averaging around \textit{16.7 clicks per second}, and significantly reduces user fatigue.

\subsection{Calibration and Performance Evaluation}

Before each session, participants complete a precision test that measures the distance of pixels from a central target across repeated taps. Additionally, they assess interaction rate using a short clicking-speed task and log runtime metrics during stimulus presentation (e.g., clicks per second and inter-click intervals). 

\subsubsection{Precision Test}

\begin{figure}[htbp]
  \centering
  \includegraphics[width=1\linewidth]{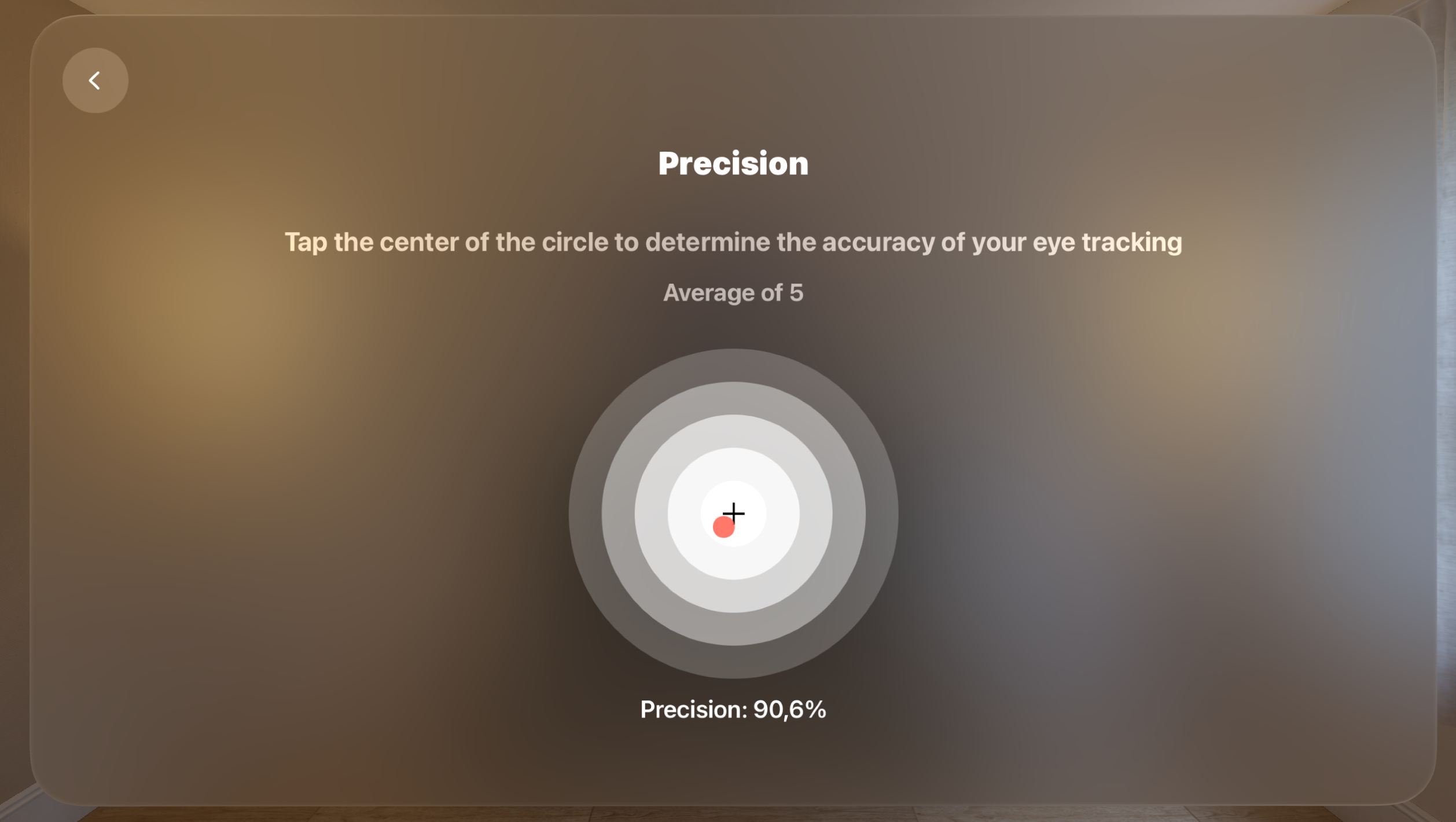}
  \caption{Precision calibration interface: users tap the center cross to measure eye‐tracking accuracy, with the red marker indicating the recorded gaze point and the resulting precision score displayed below.}
  \label{fig:precision_test}
\end{figure}

We developed a simple precision test to evaluate how accurately users have calibrated their gaze on the Apple Vision Pro. The test, illustrated in Figure~\ref{fig:precision_test}, consists of a circle with a target in the middle, where users are expected to focus their gaze and click using their preferred method. The system records the coordinates of each click and calculates their distance in pixels from the target center. Each participant has to complete five attempts for their average precision score to be evaluated as a percentage between 0 and 100. 

This metric directly relates to how accurately the system can capture the intended gaze location, with lower scores indicating lower precision. This precision score also approximates the accuracy with which the user's gaze will be captured during eye-tracking sessions. 

\subsubsection{Clicking Speed Test}

\begin{figure}[htbp]
  \centering
  \includegraphics[width=1\linewidth]{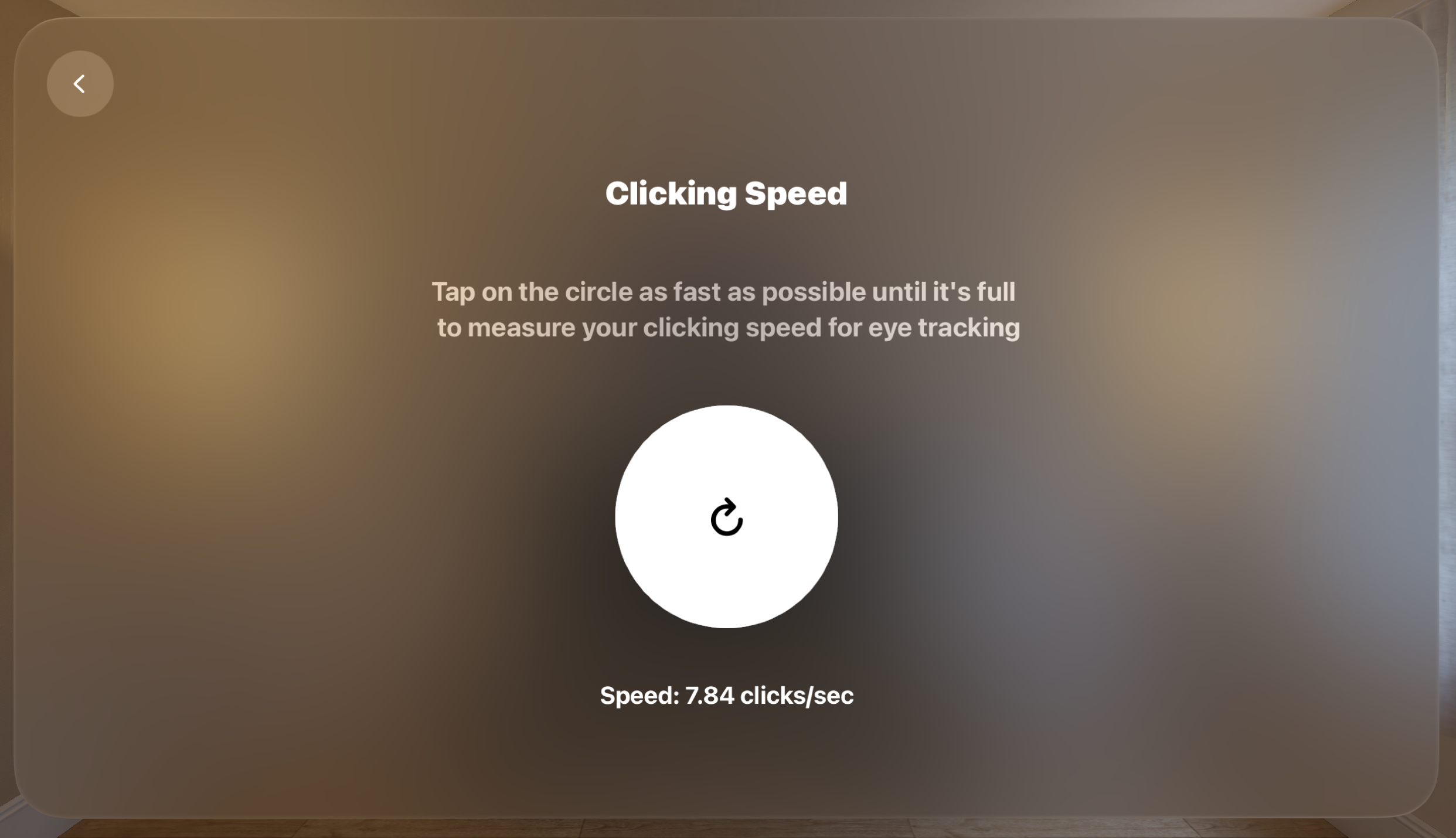}
  \caption{Clicking speed assessment interface: users tap the circle repeatedly to fill it, and the measured click rate is displayed upon completion.}
  \label{fig:clicking_speed_test}
\end{figure}

The clicking speed test assesses the interaction rate of different clicking methods. As shown in Figure~\ref{fig:clicking_speed_test}, the participants are instructed to perform 20 clicks as fast as possible on a circular target using their preferred clicking method. The system measures the total time to complete all clicks and calculates the average clicks per second. This metric provides insight into the frequency of gaze data collected during eye tracking. The higher the frequency, the denser the gaze data, allowing for a more detailed heatmap and eye tracking analysis.

\subsection{Heatmap Generation}

The heatmap generation process is a crucial aspect of the system. It involves transforming eye-tracking coordinates into a dynamic temporal heatmap, providing a visual representation of the user's gaze over time. The visualization enables detailed analysis of the user's attention, highlighting which areas of the video were in focus. 

The gaze data consists of coordinates represented as normalized floating-point numbers ranging between 0 and 1 and a timestamp marking when each point was recorded during the video. The x and y coordinates are expressed as percentages of the gaze location within the width and height of the video, respectively. This normalization approach ensures the correct coordinate mappings across different video resolutions and aspect ratios. The coordinate transformation process begins by converting the normalized x and y values to pixel coordinates within the target video dimensions. Then, a three-dimensional brightness array is created, representing the video frames along with their width and height coordinates. This array is later used to calculate the brightness of each gaze data point. 

\subsubsection{Temporal Spreading and Fade Animation}

Temporal spreading is applied to the brightness array to create smooth transitions between the heatmap frames. Each gaze coordinate is spread over multiple frames using linear interpolation to calculate the brightness values. Using a fade duration of 0.3 seconds, the brightness of each gaze point is gradually increased until it reaches its maximum at the exact timestamp of the gaze occurrence, then decreased symmetrically. This creates a fading animation to the heatmap points, making them appear smoother and avoiding sudden changes between frames. 

\subsubsection{Brightness Accumulation and Normalization}

When multiple gaze points occur at the exact pixel location, either simultaneously or with overlapping fade durations, their brightness values are additively combined. This is done to ensure that the coordinates that are examined more appear brighter than those that are examined less. A brightness normalization is applied to prevent certain areas from becoming too bright and overwhelming the visualization. The algorithm applies a square root normalization to reduce the brightness of frequently viewed points while maintaining visibility for those viewed less regularly. This ensures that all gaze coordinates are visible in the final heatmap, creating a balanced visualization.  

\subsubsection{Gaussian Blur and Color Mapping}

\begin{figure}[htbp]
  \centering
  \includegraphics[width=1\linewidth]{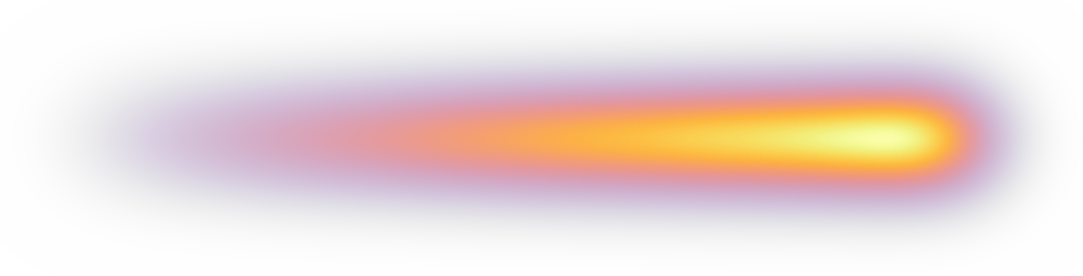}
  \caption{Heatmap of an increasing number of gaze points from left to right}
  \label{fig:heatmap}
\end{figure}

So far, the brightness array consists of sharp, isolated pixels that are not visually informative or appealing. A Gaussian blur can be applied to each frame to create a smooth circular gradient that is brighter in the center and fades toward the edges. This creates bigger regions that can be more easily interpreted. An inferno color map is applied to the brightness values to make the heatmap even more visually appealing. The mapping ensures that high-attention areas appear in warm, yellow and orange colors, while low-attention regions are represented in cool, dark purple and blue tones. Figure~\ref{fig:heatmap} shows an example of a final generated heatmap with a gradually increasing number of gaze points, where the locations on the right have been looked at more than the ones on the left. 

\subsubsection{Video Overlay}

The video overlay process combines the generated heatmap with the background video. To speed up the generation, the quality of the original video is reduced. After that, a darkened filter is applied to make the heatmap stand out more and ensure good visibility over the content. Finally, the video is added as a background to the generated heatmap. A static frame is merged to the end of each heatmap video, displaying a cumulative heatmap. This final frame combines all gaze points recorded throughout the entire video, serving as a summary of the overall attention areas.

\subsubsection{Resulting Files}

Once the overlay is complete, the resulting heatmap video is automatically saved to a folder on the device running the server. By default, this folder is called "Heatmap" and is located on the user's desktop. In addition to the video file, a JSON file is generated and stored in the exact location. This file contains the raw gaze data, including the coordinates and timestamps of each point, as well as metadata such as the user information and video name.

\subsubsection{Averaged Heatmap}

The averaged heatmap implementation aggregates the gaze information from multiple users into a single heatmap video. Before beginning the process, the video and each user's eye-tracking JSON files obtained while watching the same video should be placed in a folder. By running the Python script and including this folder path as a command-line argument, the system combines all users' gaze information into a single set, which can then be used to generate the heatmap in the same process described above. The final result is a video that contains an averaged heatmap in every frame, providing information about the collective viewing patterns of multiple users. 

\subsection{Video Eye Tracking}

\begin{figure}[htbp]
  \centering
  \begin{subfigure}[b]{0.67\linewidth}
    \centering
  \includegraphics[width=1\linewidth]{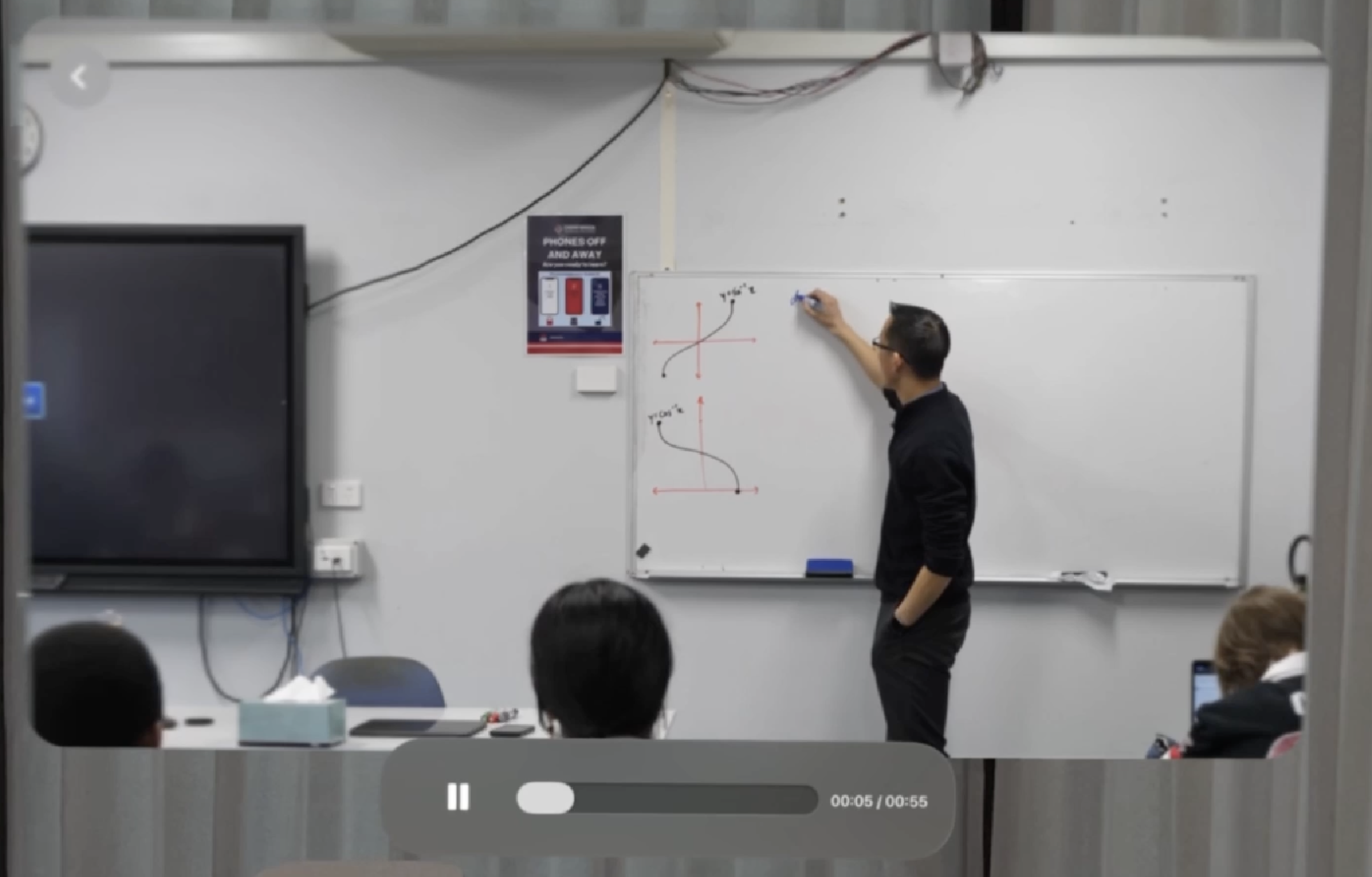}
  \end{subfigure}
  \hfill
  \begin{subfigure}[b]{0.67\linewidth}
    \centering
  \includegraphics[width=1\linewidth]{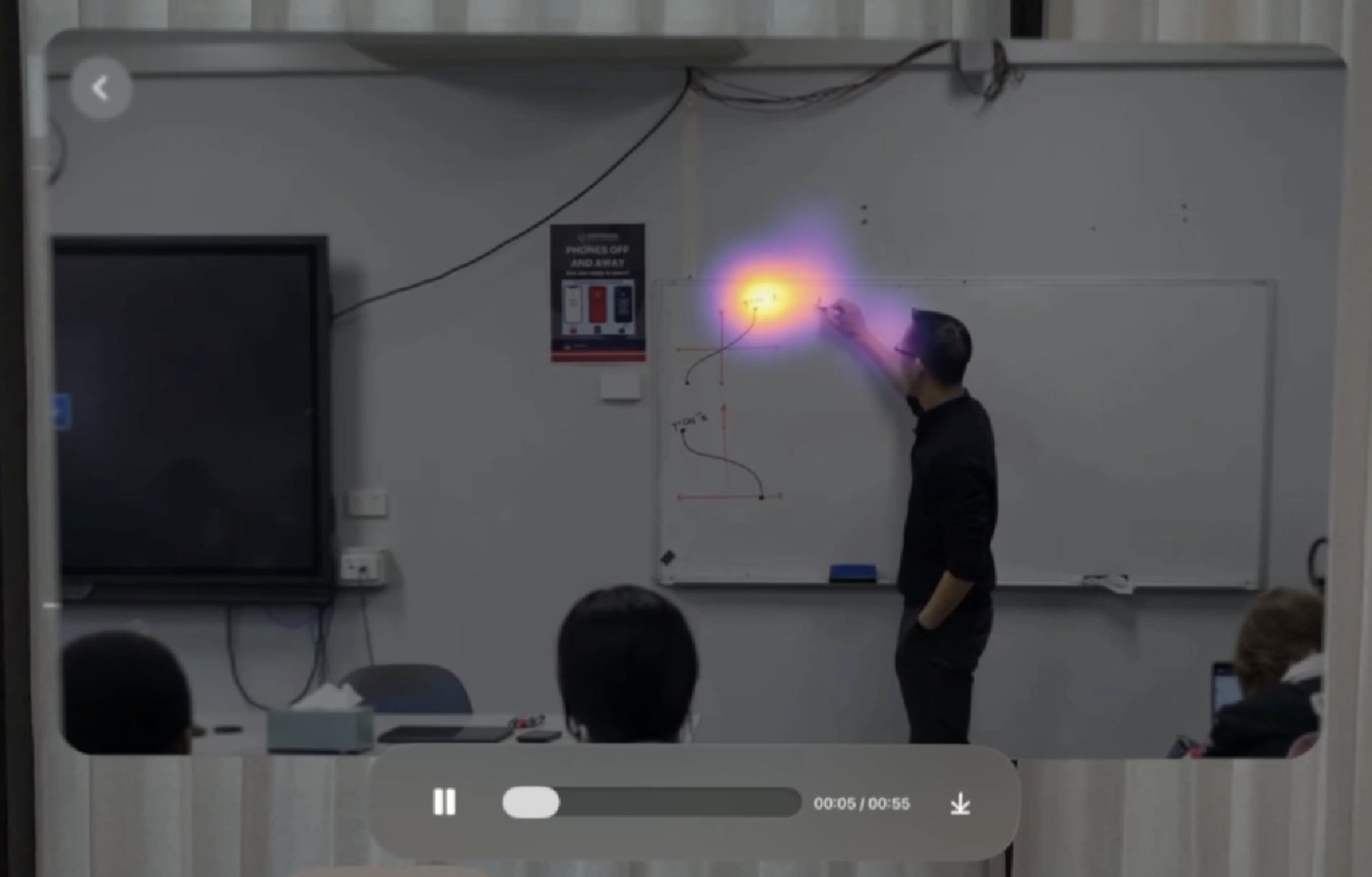}
  \end{subfigure}

  \caption{Video eye tracking interface: (top) gaze collection process during video playback and (bottom) generated heatmap visualization}
  \label{fig:video_eye_tracking}
\end{figure}

Video eye tracking refers to capturing the user's gaze movements. At the same time, they watch a pre-recorded video, and then display it as a heatmap to visualize areas of interest and attention. Figure~\ref{fig:video_eye_tracking} illustrates the video eye tracking interface and resulting heatmap visualization.

\subsubsection{Interface Design}

The video eye tracking process begins by prompting the users to upload a video. After selecting a video, the user is presented with a video player window designed to minimize distractions during eye-tracking, while still providing essential functionalities. The UI consists of a toolbar positioned outside the main video playback area, ensuring the controls do not attract unwanted attention. These controls include a play and pause button, as well as a slider for rewinding or fast-forwarding the video. This mimics a natural viewing environment that most users are already used to. A back button is also provided, allowing the user to exit the eye-tracking process. The back button must be held for a few seconds to activate, preventing accidental exits, especially when using turbo-clicking to capture gaze data. This UI was developed following the guidelines for best practices in \cite{mehmedova2025virtual}

\subsubsection{Gaze Capture}
A transparent rectangle is placed on top of the video to capture the exact position of the user's gaze when a click event occurs. On this new layer, an onTapGesture click event listener is used to detect the location of the user's gaze when a click is performed. The obtained coordinate points are then translated to percentages of the screen's width and height to ensure they are independent of the video resolution. Each coordinate is also associated with a timestamp of when the click occurred within the video’s timeline.

\subsubsection{Data Transfer and Processing}
When the video is over, the collected gaze data, video name, and user information, including their precision scores, are combined in a JSON file. Using an HTTP \verb|POST| request, the video and the JSON file are then sent to the server, where the heatmap video can be generated. Once the generation is complete, the resulting video is returned to the Swift application via an HTTP response. 

\subsubsection{Background Processing and Export}
If the user does not want to wait for the heatmap generation, they are given the option to run the process in the background. In this case, once completed, the heatmap video and its corresponding JSON file are saved only to the server device. Alternatively, if the user chooses to wait, the heatmap video will be displayed once generated. The user can then export the video and JSON file to the Vision Pro device. 

\subsection{Spatial Eye Tracking}

\begin{figure}[htbp]
  \centering
  \begin{subfigure}[b]{0.9\linewidth}
    \centering
  \includegraphics[width=1\linewidth]{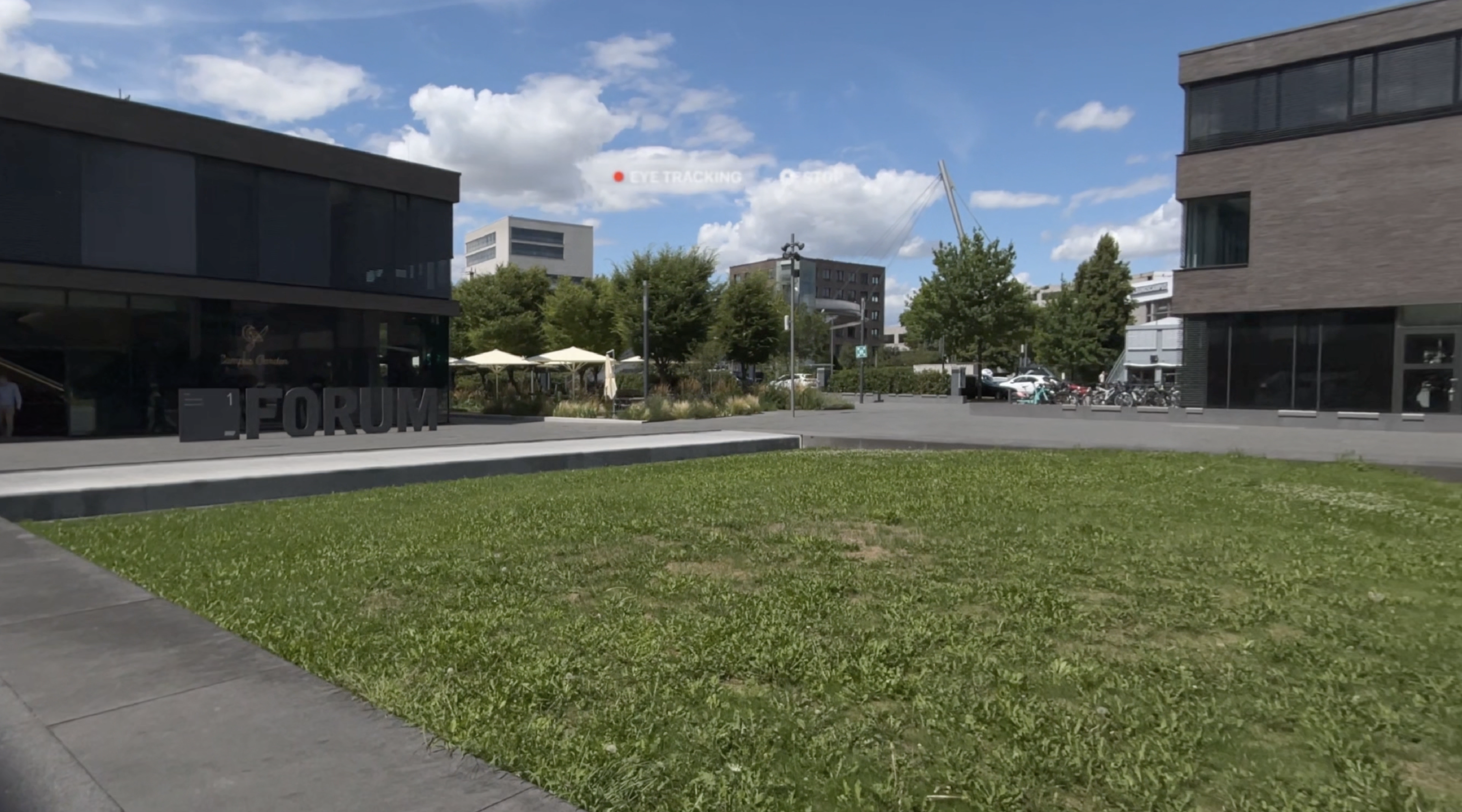}
  \end{subfigure}
  \hfill
  \begin{subfigure}[b]{0.9\linewidth}
    \centering
  \includegraphics[width=1\linewidth]{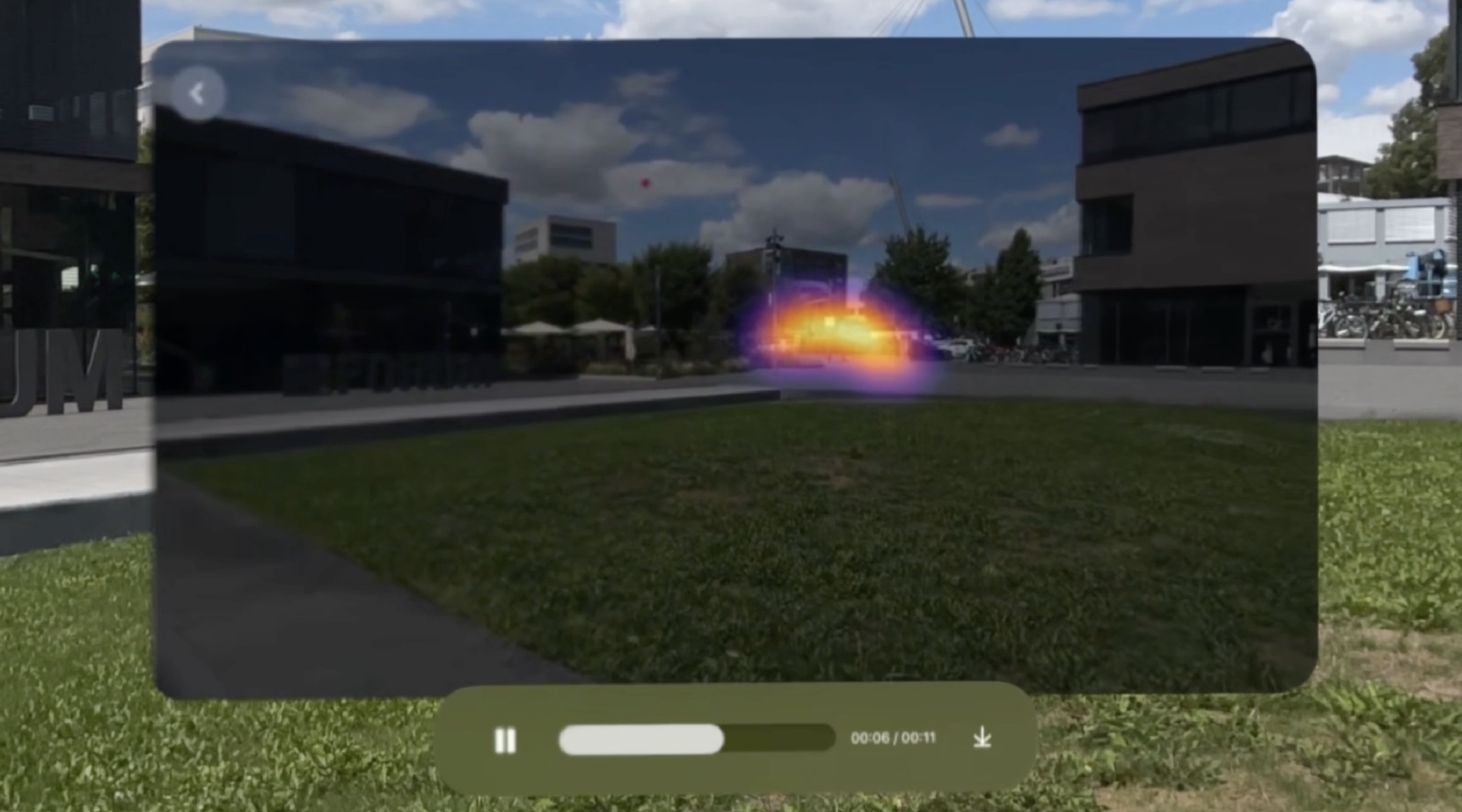}
  \end{subfigure}

  \caption{Spatial eye tracking interface: (top) spatial gaze collection process and (bottom) generated heatmap visualization.}
  \label{fig:spatial_eye_tracking}
\end{figure}

Spatial eye tracking extends traditional video-based gaze analysis by capturing users' eye movements in real-world environments. Unlike video eye tracking, which analyzes gaze on prerecorded media, spatial eye tracking requires simultaneous recording of the user's physical surroundings and gaze coordinates. This provides insights into how users interact visually with their environment. The spatial eye tracking implementation is demonstrated in Figure~\ref{fig:spatial_eye_tracking}.

\subsubsection{Environmental Recording Challenges}

The most crucial part of spatial eye tracking is obtaining a video of the user's surroundings while they look around. The Vision Pro provides a native video recording functionality of the user's space; however, accessing this feature requires a special entitlement. These entitlements are designed to protect the user's privacy and ensure that certain features are used responsibly. The entitlement application process requires official approval from Apple, but the timeline for obtaining this approval is uncertain. Given the time constraints of this research project and the uncertainty of the approval outcome, an alternative approach was necessary to fulfill the spatial video recording functionality.

\subsubsection{Screen Mirroring Approach}

A workaround solution focuses on leveraging existing Vision Pro capabilities to capture a video of the user's visual field without requiring additional entitlements. The view mirroring functionality provided by visionOS offers a viable alternative for capturing the user's environment while maintaining the ability to collect gaze coordinates simultaneously. Once the Vision Pro display is shared with a Mac device, it can be screen recorded to obtain the spatial video. 

When spatial eye-tracking is initiated, the Swift application sends a request to the server to begin screen recording. Simultaneously, the user's gaze coordinates are recorded locally in the application. Once the session is stopped, a JSON file containing the gaze data, along with a request to stop recording, is transmitted to the server. The server then generates a heatmap and combines it with the recorded footage to produce the final heatmap video, which is returned to the Vision Pro application for display to users. As with the video-based eye tracking, users can wait for the heatmap to be generated or choose to run the process in the background. 

\subsubsection{Implementation Details}

The spatial eye tracking implementation is designed with attention to user experience, ensuring users remain focused on their surroundings without being distracted by interface elements. An immersive space was chosen for this purpose, as it enables the placement of elements in 3D space without obstructing the user's entire field of view. Only essential interface elements are visible within this space, including a recording indicator and a stop button, both of which are positioned at the top of the view to minimize visual distraction.

A transparent rectangle is placed in front of the user's view to record the gaze coordinates in this immersive space. Similar to video eye tracking, a click event listener is used to obtain the exact gaze location points within the rectangle, expressed as a percentage of the overall width and height. Each coordinate is also associated with a timestamp relative to the start of the spatial tracking.

The size and positioning of the transparent rectangle are crucial for accurate eye tracking. If it doesn’t align perfectly with the portion of the user's view being mirrored to the Mac device, the gaze coordinates will be mapped incorrectly, resulting in a heatmap that does not accurately reflect where the user was looking. To ensure the transparent rectangle aligns perfectly with the area of the mirrored portion of the view, a visible border was temporarily added to the rectangle. This made it easier to manually adjust its position and size. Another adjustment that needed to be made was cropping the screen video recording, as black borders appeared on top and bottom of the mirrored view.

To verify that the coordinates are mapped correctly, a dot was temporarily displayed on the rectangle at each gaze location. The server then used those exact coordinates to generate a heatmap frame with the same-sized dots, and the results were compared to ensure the server-rendered dots appeared in the same position as the original ones. After several rounds of fine-tuning and adjustments, the alignment was calibrated to reproduce the accurate gaze positions.

\section{Experiment}\label{E}

An experiment evaluated the effectiveness of the developed eye-tracking system and compared different interaction methods. The study employs a between-subjects design, with two distinct groups, each utilizing different click methods to collect gaze data. 

\subsection{Participants}

We recruited 20 participants and randomly assigned them to two equal groups (n = 10 per group). 

\textit{Inclusion.} Adults (18+) able to complete Apple Vision Pro calibration and the study tasks; normal or corrected-to-normal vision with contact lenses permitted.

\textit{Exclusion.} Individuals who wear prescription eyeglasses were excluded because, in our setup, the Apple Vision Pro could not be comfortably or safely used with glasses, and compatible optical inserts were not available for our study sessions. Participants with known visual, neurological, or motor conditions that could affect eye tracking or input performance were also excluded.

Each group was in a between-subjects design comparing two methods for obtaining click-gated gaze information on the Apple Vision Pro: dwell versus game controller input. The pinch gesture was excluded as a clicking method due to user fatigue reported during pilot testing.
Before starting the experiment, each participant completed the Vision Pro’s built-in eye-tracking calibration to optimize the system for their individual gaze characteristics. Immediately after calibration, participants performed a precision test to assess the accuracy of their gaze setup, ensuring that subsequent recordings accurately reflected the calibration.

\textbf{Dwell Control Group.} 
This group of participants utilized the Apple Vision Pro's built-in dwell control functionality as their primary way of clicking for gaze collection. To allow the fastest response time, it was configured for the fastest clicking speed and the highest movement tolerance. This interaction method allowed participants to register clicks hands-free.

\textbf{Gaming Controller Group.}
This group used an 8BitDo Pro 2 gaming controller connected to the Vision Pro via Bluetooth. The controller was configured to enable turbo clicking, allowing users to perform fast clicks on a single button press.  

All participants provided written informed consent prior to participation and were free to withdraw at any time without penalty. No personally identifying information was collected.

\subsection{Eye Tracking Content}

Two videos were selected to assess gaze patterns across different content types.  

\textbf{Lecture Video.} 
A short educational mathematics lecture video\footnote{Youtube video url: https://www.youtube.com/watch?v=gTSTrCJ0LDg} was selected to assess the visual attention during instructional content. This video features a lecturer presenting a concept while being interrupted by distractions from other students and background elements. 

\textbf{Quiz Questions.}
A video containing two Unified Modeling Language (UML) quiz questions, displayed as static images, one after the other. Each question consisted of a diagram, a question, and four possible answer choices. This type of content was used to evaluate the participants' gaze patterns during problem-solving tasks and visual information processing.

\section{Results}\label{R}

The findings of our user study focus on calibration precision, click-based data collection rates, and the visual output of generated heat maps. We compare the performance of dwell control versus gaming controller interactions and illustrate how these methods influence the density and accuracy of gaze visualizations.

\subsection{Precision Scores}

\begin{figure}[htbp]
  \centering
  \includegraphics[width=0.95\linewidth]{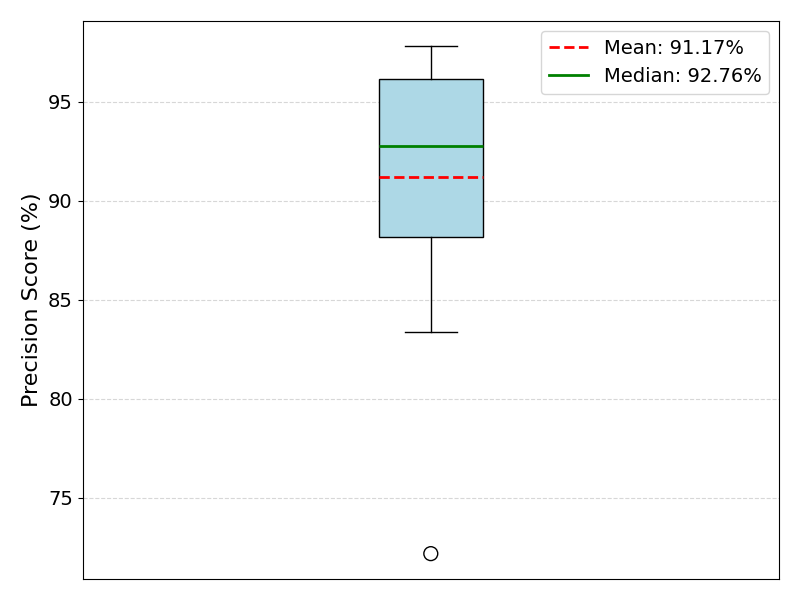}
  \caption{Average precision score}
  \label{fig:precision}
\end{figure}

The precision scores of each user represent how accurately they have set up the eye tracking on the Apple Vision Pro device. This score also approximates the precise gaze collection of each participant. Figure~\ref{fig:precision} depicts the precision scores across all participants, ranging from approximately 73.7\% to 97.8\%, with most values clustered above 85\%. The median precision score is 92\%, indicating that at least half of the participants achieved high calibration accuracy. 

The mean score is slightly lower at 91\%, reflecting a slight influence from a single participant who achieved a lower score. Overall, the plot suggests that most users could calibrate their eye tracking with good precision, making the collected gaze data reliable for further analysis.

\subsection{Gaze Data Collection Frequency}

To evaluate the effectiveness of each clicking method, the collected gaze data was analyzed and compared between the two videos.

\subsubsection{Total Clicks per Video}

Table~\ref{tab:click_counts} presents the average total clicks collected while watching the full lecture and quiz videos from the 10 participants who used the gaming controller and the other 10 participants who watched them with dwell control. The gaming controller gaze collection method produced over 30 times more clicks than dwell control across the lecture and quiz videos. This highlights the big difference in data collection frequency between the two interaction techniques. The higher number of clicks in the quiz video can be explained by its longer duration, which naturally allows for more gaze interactions over time.

\begin{table}[htbp]
\centering
\caption{Average Click Count by Method and Video Type}
\label{tab:click_counts}
\begin{tabular}{@{}lcc@{}}
\toprule
\textbf{Video Type} & \textbf{Dwell Control} & \textbf{Gaming Controller} \\
\midrule
Lecture & 25.4 & 784.3 \\
Quiz    & 41.1 & 1307.7 \\
\bottomrule
\end{tabular}
\end{table}

\subsubsection{Average Clicks per Second}

\begin{figure}[htbp]
  \centering
  \includegraphics[width=1\linewidth]{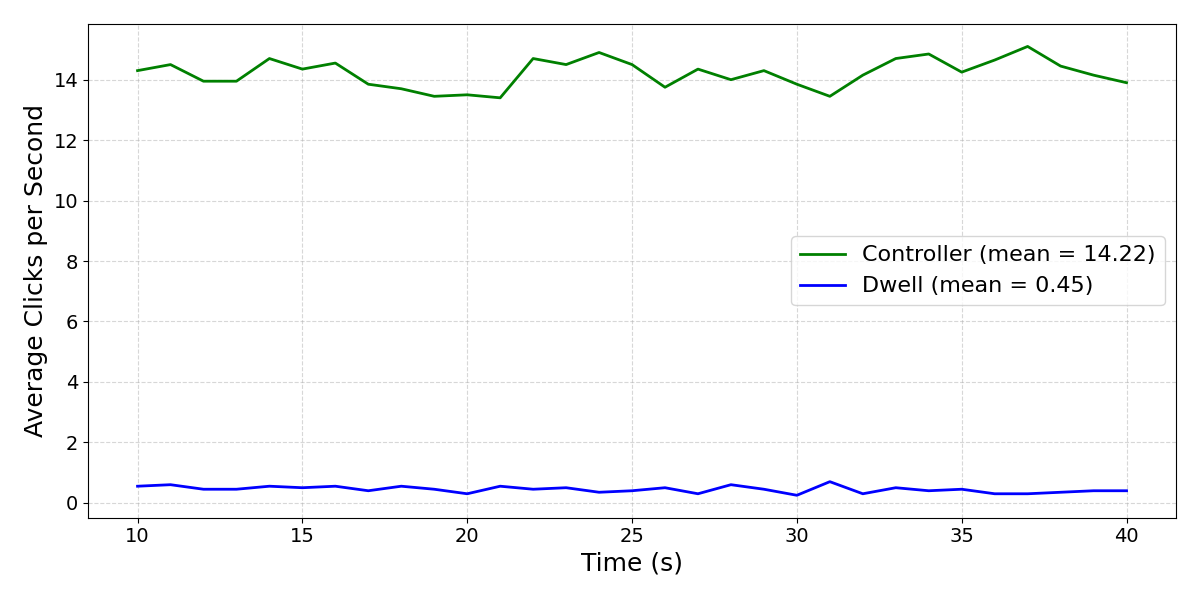}
  \caption{Average clicks per second}
  \label{fig:clicks_per_sec}
\end{figure}

The frequency of the clicks becomes more apparent when we examine the average number of clicks per second. Each value in the line chart shown in Figure~\ref{fig:clicks_per_sec} is an average score of the clicks per second of the participants who used a controller and those who used dwell control. Data from both the lecture and quiz videos was used to generate the results, and display the average clicks for each second of the video. The time window selected for this analysis spans a 30-second interval, from the 10th to the 40th second, of both videos. 

This specific range was chosen to capture the period of consistent gaze interaction and exclude the initial moments of playback when participants might have been adjusting to the task and not yet interacting with clicks. The results show a difference between the two interaction methods. The controller group achieved a much higher mean click rate of 14.22 clicks/s, while the dwell group averaged only 0.45 clicks/s. 

It is worth noting that in a separate self-conducted clicking speed test, where the average speed of the controller and dwell controller was measured as an average of 10 attempts, the results were slightly higher for both methods. The average speed for the controller was 16.78 clicks per second (clicks/s), while dwell-controlled achieved 0.73 clicks/s. The difference in speeds observed in this experiment is due to the varying methods used to evaluate clicking speeds. When measured using the clicking speed test, the target being observed was a stationary object, and the test did not account for different gaze movements. In the current experiment, however, participants frequently shifted their gaze around the video, making it harder for dwell control, in particular, to trigger clicks consistently due to its movement tolerance threshold. 

\subsubsection{Average Inner Click Interval}

To further evaluate the frequency of the gaze data, the average inner click interval is calculated for both interaction techniques. This metric represents the average time between successive click events. 

\begin{figure}[htbp]
  \centering
  \includegraphics[width=1\linewidth]{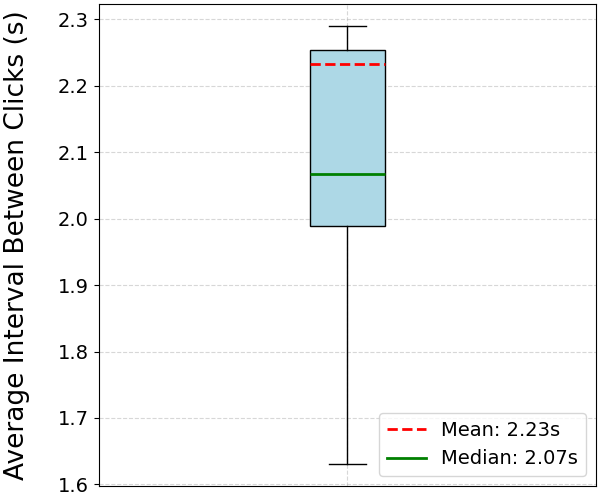}
  \caption{Average inner click interval for dwell control.}
  \label{fig:inner_dwell}
\end{figure}

As expected, the dwell-based interaction method shown in Figure~\ref{fig:inner_dwell} resulted in significantly longer intervals between clicks, averaging around 2.23 seconds, while the median was 2.07 seconds. This suggests there is a significant gap between the collected gaze points. The box plot also shows a relatively wide spread, indicating a higher variability in dwell clicking, likely caused by the different gaze movements of the participants. 

\begin{figure}[htbp]
  \centering
  \includegraphics[width=1\linewidth]{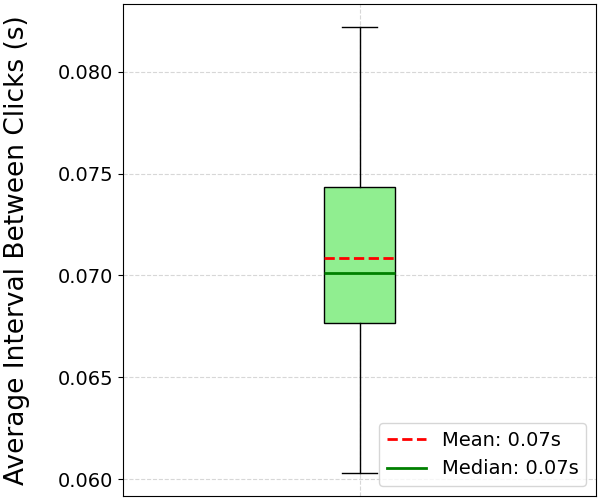}
  \caption{Average inner click interval for the gaming controller.}
  \label{fig:inner_controller}
\end{figure}

In contrast, the controller-based approach illustrated in Figure~\ref{fig:inner_controller} averaged 0.7 seconds between clicks and had a narrower distribution. This again confirms that a higher click frequency can be achieved when using a gaming controller to collect gaze points.

\subsection{Heatmap Videos}

Collected coordinates are mapped from normalized space to video pixels and accumulated over time into a brightness volume. Then, it is possible to render two types of heatmaps. 

\subsubsection{Individual Heatmap Videos}

Individual heatmap videos were created for each participant, who watched the two videos, resulting in 40 visualizations. These heatmaps visually represent the participant's individual attention patterns, revealing unique viewing behaviors and focus areas specific to each user. As expected, the heatmap videos of the group that used a gaming controller to collect their gaze coordinates were much denser than those of the group that used dwell control. The controller enabled continuous eye-tracking, whereas dwell control provided individual gaze points at intervals of a few seconds.   

\subsubsection{Averaged Heatmap Videos}
An averaged heatmap approach was used to visualize the collective viewing patterns. The resulting videos are an aggregation of all participants' gaze data within each group. A total of four averaged heatmap videos were created: the lecture video with dwell control, the lecture video with the controller, the quiz video with dwell control, and the quiz video with the controller. 

\begin{figure}[htbp]
  \centering
  \includegraphics[width=1\linewidth]{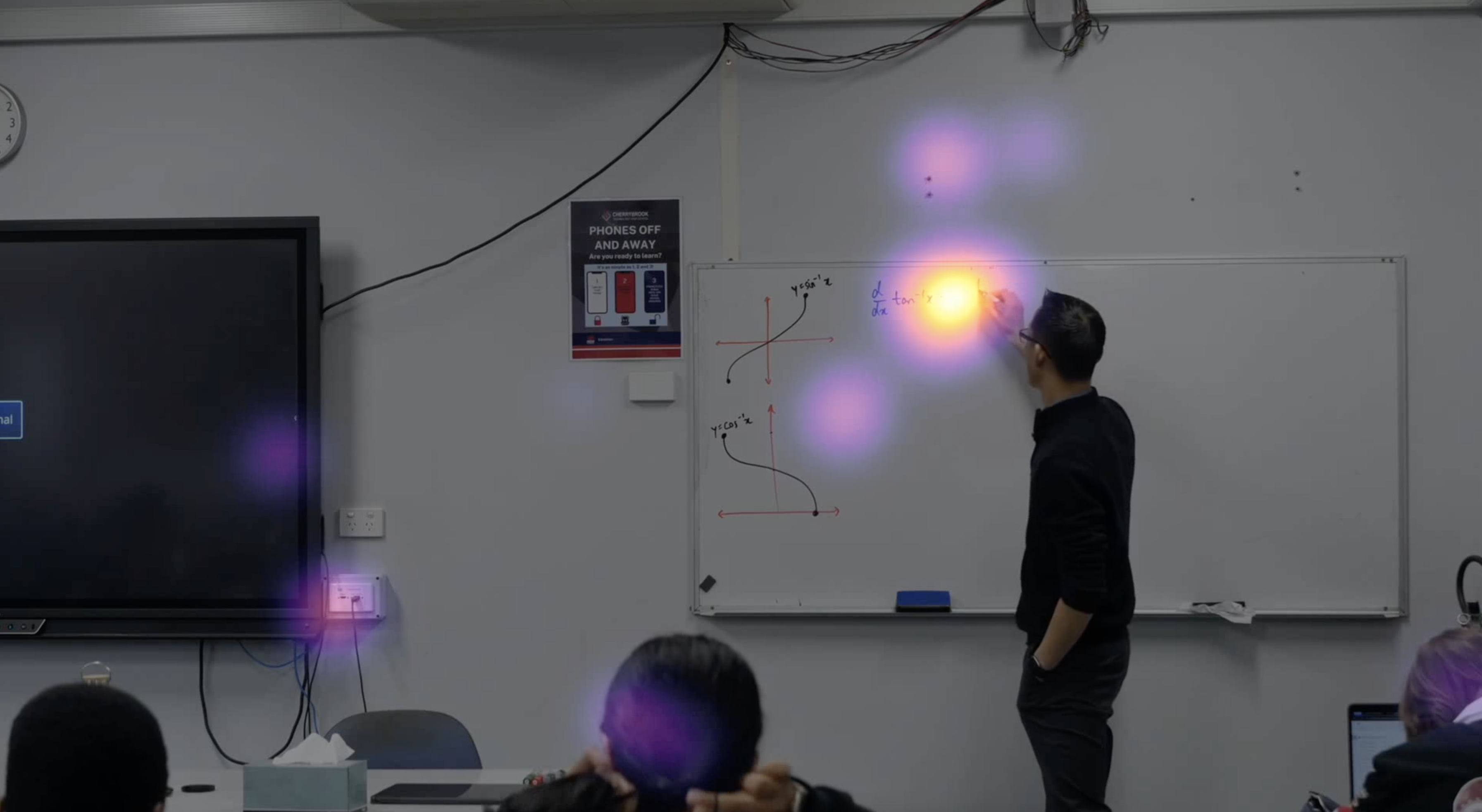}
  \caption{Averaged heatmap frame from the lecture video obtained by using a gaming controller}
  \label{fig:attention}
\end{figure}

As shown in Figure~\ref{fig:attention}, the averaged heatmap videos obtained from using the controller provide valuable information on where the majority of the participants were focused in each frame of the video, while also illustrating further areas of attention, where other participants were looking. 

\begin{figure}[htbp]
  \centering
  \begin{subfigure}[b]{0.492\linewidth}
    \centering
  \includegraphics[width=1\linewidth]{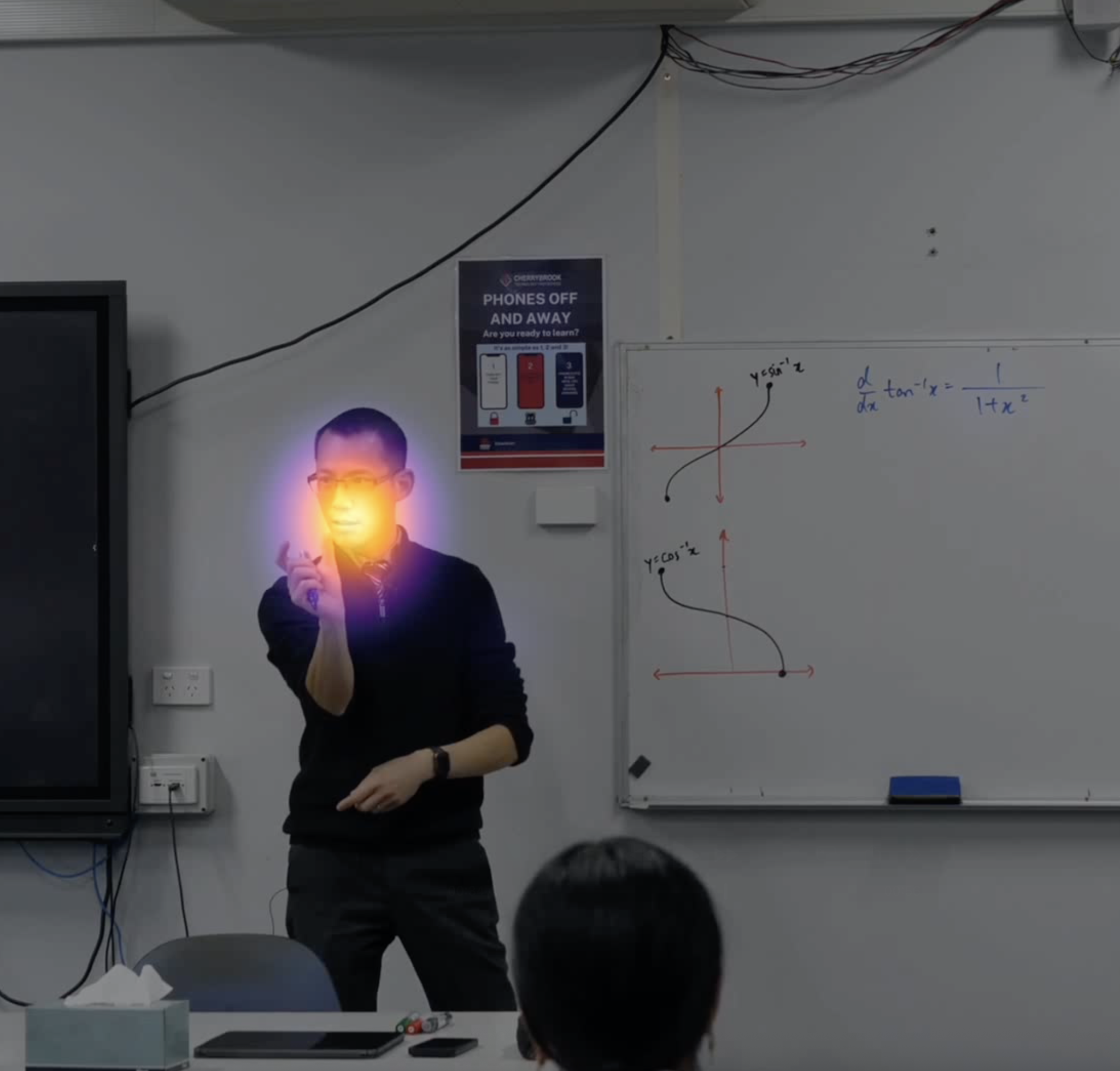}
  \end{subfigure}
  \hfill
  \begin{subfigure}[b]{0.492\linewidth}
    \centering
  \includegraphics[width=1\linewidth]{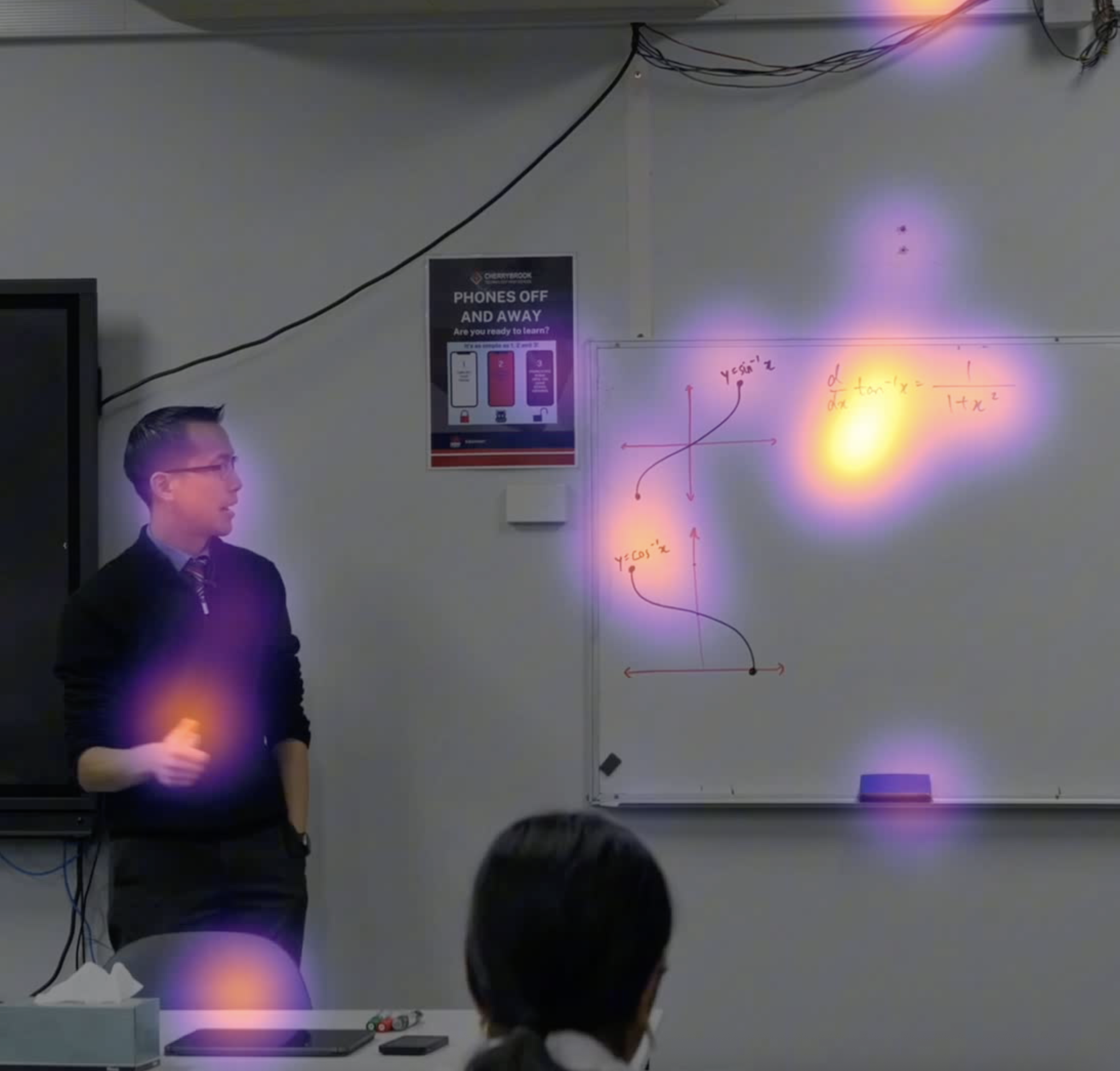}
  \end{subfigure}

  \caption{Averaged heatmap video frames of (left) clustered and (right) scattered attention}
  \label{fig:focus}
\end{figure}

Moreover, frames with a single, clustered heatmap region, as demonstrated in Figure~\ref{fig:focus}, suggest a shared focus among the participants, whereas frames with multiple scattered gaze points imply divided attention and a higher likelihood of distraction. 

\begin{figure}[htbp]
  \centering
  \includegraphics[width=1\linewidth]{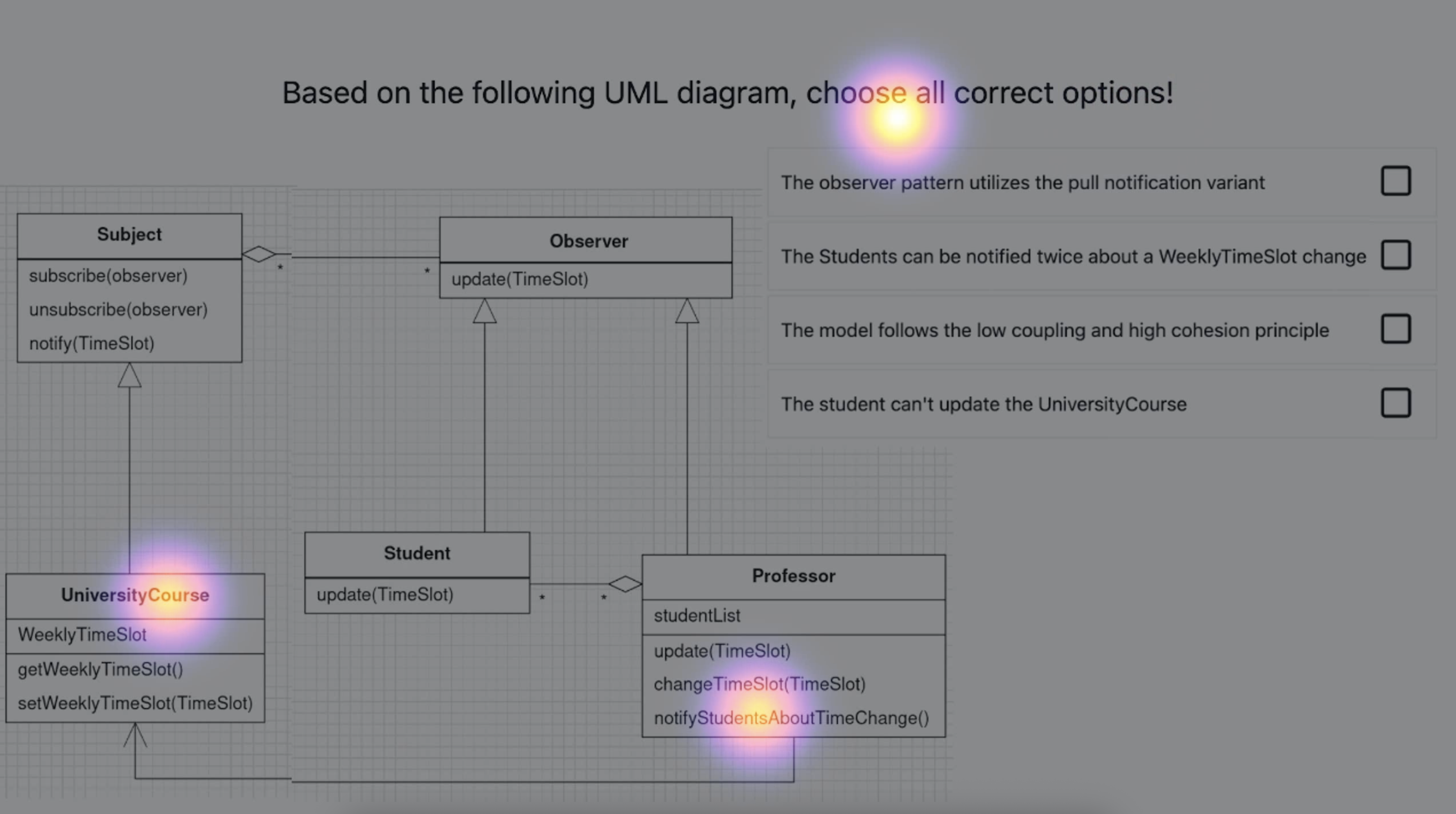}
  \caption{Averaged heatmap frame from the quiz video obtained by using dwell control}
  \label{fig:dwell_heatmap}
\end{figure}

As illustrated in Figure~\ref{fig:dwell_heatmap}, the averaged heatmap videos obtained from the group that utilized dwell control to click provide more scarce but still valuable information about the main areas of focus. If performed with a bigger group size, this method can provide an even denser and more accurate averaged heatmap. 

\begin{figure}[htbp]
  \centering
  \includegraphics[width=1\linewidth]{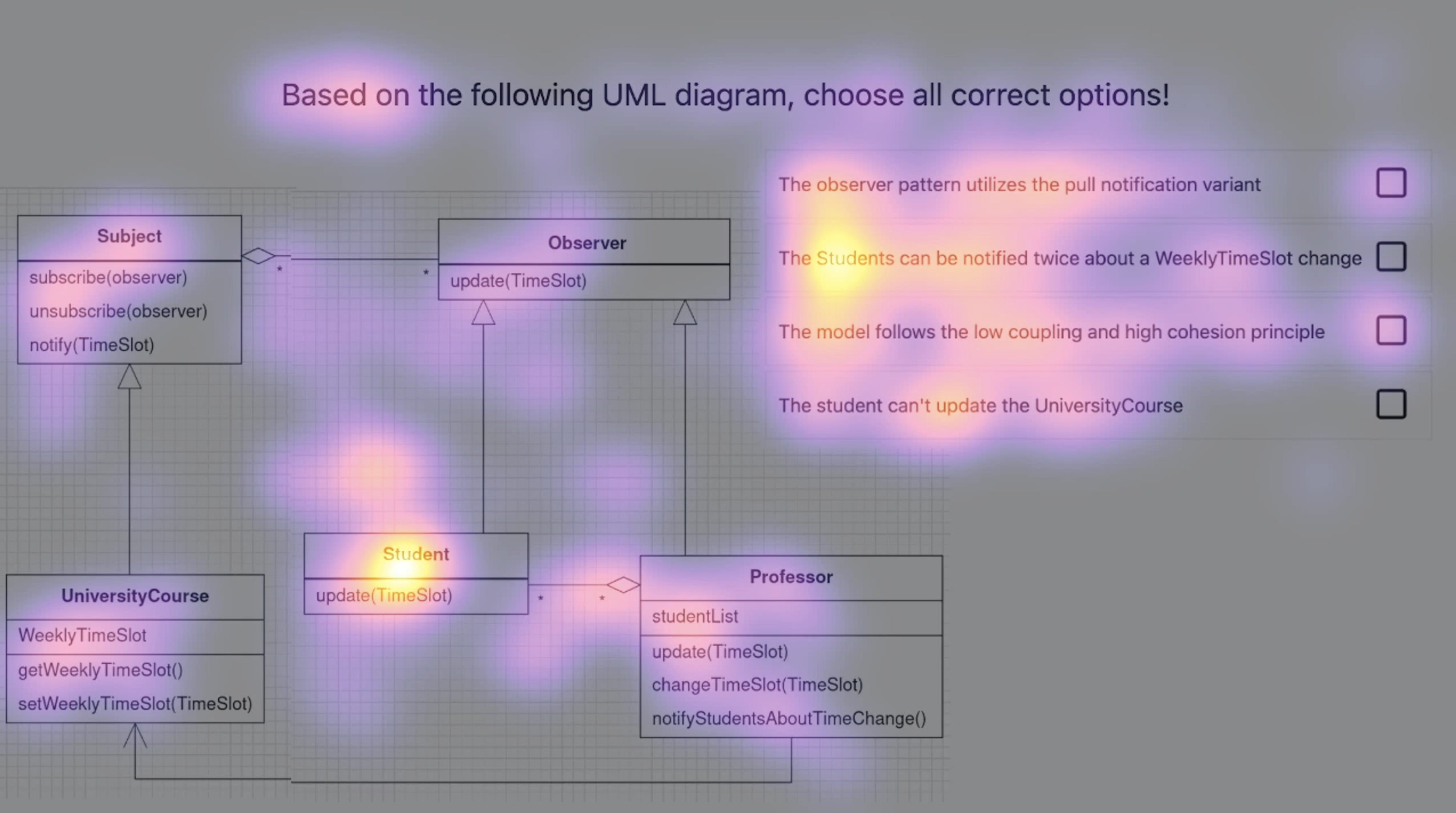}
  \caption{Final heatmap frame of the averaged dwell control quiz video, displaying all gaze points from all participants at the same time}
  \label{fig:final}
\end{figure}

The final heatmap frame, as depicted in Figure~\ref{fig:final}, comprises the gaze data of all participants across all frames, effectively highlighting the regions that consistently attract their attention.

\section{Discussion}\label{D}

This work presents a novel system for collecting and visualizing gaze data on the Apple Vision Pro. Utilizing client-server communication allows the creation of dynamic heatmaps across videos and spatial environments. The system also provides a tool for generating averaged heatmaps, which can help evaluate the collective gaze attention of a group. Moreover, the study proposes several options for extracting gaze coordinates and evaluates these methods in a user experiment. 

\subsection{Applications}

The flexibility of using a mixed reality headset for eye tracking purposes enables a wide range of possible applications. The video-based and spatial gaze tracking functionalities on the Apple Vision Pro enable researchers to effectively visualize and analyze the user's gaze across various domains, including education, design, marketing, and cognitive science.

\subsubsection{Education}

In the education field, the system has the potential to benefit both researchers and instructors by revealing how students visually interact with learning materials and lectures. Video eye tracking can accurately assess how students approach problem-solving tasks or engage with online lectures and slides. On the other hand, spatial eye tracking can be used for live in-person lectures to provide valuable insight into when the students are focused on the lecture and what causes disengagement and distraction. 

\subsubsection{Environmental Analysis}

Spatial eye tracking can be used in interior and exterior environments to study how people interact and move through physical places. In museums, galleries, and exhibitions, the gaze data can reveal which artworks or exhibits attract people's attention more, and how different layouts affect their engagement. In architecture and interior design, eye tracking can help identify how different design choices influence the user's attention and what they focus on most. It can also provide crucial information on navigating a new building or paying attention to safety signs. This can help architects and designers optimize spatial layout, wayfinding, and visual comfort in physical places. 

\subsubsection{Marketing and Advertising}

In marketing and consumer behaviour research, this tool can be applied to study visual attention in digital and physical advertising. In video advertisements, researchers can study which products, logos, and faces receive more attention and which are ignored. In physical retail environments and product demonstrations, analyzing the customer's gaze location and duration can help improve product placement, shelf layout, and packaging design. 

\subsubsection{Clinical Science}

The system also has applications in clinical psychology and cognitive science. Eye tracking can be used to study visual attention patterns in patients with different conditions, such as autism, ADHD, or traumatic brain injuries. Researchers can compare gaze behavior across multiple age groups or demographic backgrounds to assess differences in attention or social engagement. In therapy sessions, spatial eye tracking could also assess focus during different treatments, such as exposure-based ones.

\subsection{Limitations}

While the system has good functionalities that can be applied, several limitations must also be acknowledged. 

\textbf{Lack of Continuous Gaze Access.}
Due to Apple's privacy restrictions, access to the continuous raw gaze data on the device is forbidden, limiting the frequency of the collected eye-tracking data. Instead, gaze coordinates are only available to the system at the moment of user interaction, such as a click or tap, which only provides discrete gaze points and limits the completeness of the generated heatmap.

\textbf{Clicking Methods.}
Depending on the click method, gaze collection also comes with trade-offs. When pinching with fingers, users are more likely to experience user fatigue in prolonged periods, or when using a gaming controller, they need to remember to hold a button. Dwell control eliminates the user fatigue issues; however, it produces fewer clicks per second and displays a visual dot when the click is performed, which can be distracting for some users.

\textbf{Spatial Eye Tracking Delay.}
Due to a lack of an entitlement, there is a slight mismatch between the gaze data and the video recording for the spatial eye tracking. When the eye-tracking process starts, a request must be sent to the server before the screen recording can begin. This can introduce a slight delay, causing gaze points to be shifted in time and not represented in the heatmap at the exact moment they occurred. 

Because both the gaze collection and video recording processes need to be synchronized as well as possible, an average delay duration was taken into account and added to the gaze collection process to better align the start times of both recordings. This does not entirely resolve the issue, as the request time to start recording may vary; however, it helps reduce the delay and better aligns the gaze data with the video. 

\textbf{Precision Test Accuracy.}
The precision test, which helps determine how accurately users have positioned their gaze on the device, is implemented on a floating window whose size and position can be adjusted. This means that if the user brings the window closer to them or increases its size, they are more likely to achieve a better precision score, which will not necessarily mean they calibrated their gaze better. That said, when a user manipulates the window, they usually keep it in the same position and size during the eye-tracking session. This means the score is still relevant to their gaze accuracy during video eye tracking.  

\section{Future Work}\label{FW}

The system can be improved and expanded in several ways to enhance its accuracy and broaden its scope of application.

\textbf{Synchronization between Gaze and video.}
An entailment can be obtained to eliminate the delay between the video recording and gaze data in spatial eye tracking by allowing the video recording on the Vision Pro device, rather than by the server. This will enable both processes to start simultaneously, and the resulting heatmap will visualize the user's gaze with more accurate timing.

\textbf{Precision score evaluation.}
The precision score can be improved to maintain a fixed distance and position for all participants, ensuring consistent initial testing conditions across users. Further improvements could include adding multiple targets to test the accuracy in different parts of the user's field of view, rather than just at a central location.

\textbf{3D Spatial heatmaps.}
The gaze data captured in the spatial eye tracking is visualized as a 2D heatmap projected onto the video. Future development can add depth to the gaze points and visualize them as a 3D heatmap. This will provide a more accurate depiction than previous works, such as \cite{rolff2022gaze}. Users will explore the space and gain a more immersive visualization, and researchers will obtain better information for in-depth analysis. 

\textbf{VR Compatibility.}
A new feature can be explored to enable eye tracking to work across other applications, websites, and virtual environments. This will enhance the system's applicability by making it more helpful for analyzing attention in various applications on the Vision Pro. Designers could track user experience and usability in websites, app prototypes, virtual environments, and games.

\section{Ethical Considerations}\label{sec:ethics}

Due to the privacy implications of eye tracking \cite{kroger2020does, gobel2020gaze, kourtesis2024comprehensive}, the \textit{iTrace} system is intended for controlled research settings only. It must be used in full compliance with Institutional Review Board (IRB) protocols and data privacy regulations. We encourage researchers to obtain informed consent from all participants, clearly explaining how their gaze data will be collected, processed, and stored. Under no circumstances should \textit{iTrace} be deployed for covert surveillance or commercial tracking of individuals without their explicit permission. All recorded gaze data should be anonymized and retained only as long as necessary for the approved study.  

\section{Conclusion}\label{C}

This research presents a practical solution for collecting and visualizing eye-tracking data on the Apple Vision Pro. The developed \textit{iTrace} application obtains gaze coordinates from user interaction events, such as pinch gestures, dwell controls, or gaming controller clicks, and transforms them into a dynamic heatmap that displays the user's gaze location while watching a video or navigating around. 
A user experiment involving 20 participants was conducted to test the system's functionality and generate averaged heatmaps from the participants' responses. It also provided measurements about the precision scores of the users and the gaze collection frequencies of the different clicking methods. While limitations exist due to Apple's privacy policies, this work establishes a foundation for eye-tracking research on the Apple Vision Pro and opens endless possibilities for further research applications.

\backmatter

\bmhead{Acknowledgements}

This research was financially supported by the TUM Campus Heilbronn \textit{Incentive Fund 2024} of the Technical University of Munich, TUM Campus Heilbronn. We gratefully acknowledge their support, which provided the essential resources and opportunities to conduct this study.

\section*{Declarations}

\bmhead{Funding} 
This research was financially supported by the TUM Campus Heilbronn \textit{Incentive Fund 2024} of the Technical University of Munich, TUM Campus Heilbronn. We gratefully acknowledge their support, which provided the essential resources and opportunities to conduct this study. 

\bmhead{Conflict of interest}
The authors declare there is no conflict of interest. 

\bmhead{Consent to participate}
The authors declare that all participants in this study consented to their participation and no personal data were collected or stored from them during the study. 

\bmhead{Data availability}
No apply. No personal data or gaze data associated with any anonymous participant is available in this study. 

\bmhead{Materials availability}
No apply 

\bmhead{Code availability}
The \textit{iTrace} source code used in this study, including the visionOS Swift client application and the Python/Flask server for heatmap generation, is publicly available at \url{https://github.com/TUM-HN/iTrace}. The repository contains build and usage instructions and scripts to reproduce the heatmap videos and averaged heatmaps reported in the paper. Exact build/runtime requirements are documented in the README (e.g., macOS/Xcode for the Vision Pro client and Python 3.x with NumPy, OpenCV, Matplotlib, and Flask for the server). 

\bmhead{Author contribution}
EM: writing, conducting the experiment. 
SBG: writing, supervision, experiment.
SW: Writing, revision, and supervision

\bibliography{sn-article}  

\end{document}